\documentclass[12pt]{article}
 \usepackage{epsfig}
 \usepackage{pstricks}

\newcommand{\dd}{\mbox{d}}
\newcommand{\ampl}[2]{{\cal M}\left( #1 \to #2 \right)}
\newcommand{\scr}{\scriptscriptstyle}

\newcommand{\sn}{\widetilde{N_j^c}}
\newcommand{\snj}{\widetilde{N_j^c}}

\newcommand{\Ynone}{Y_{\scr N_1}}

\newcommand{\Ynoneeq}{Y_{\scr N_1}^{\rm eq}}

\newcommand{\Ypone}{Y_{1+}}

\newcommand{\Ymone}{Y_{1-}}

\newcommand{\YL}{Y_{\scr L_f}}
\newcommand{\YLt}{Y_{\scr L_s}}

\def\a{\alpha}
\def\b{\beta}

\def\d{\delta}
\def\e{\epsilon}

\def\g{\gamma}

\def\j{\psi}
\def\k{\kappa}
\def\l{\lambda}
\def\m{\mu}
\def\n{\nu}

\def\p{\pi}

\def\r{\rho}
\def\s{\sigma}
\def\t{\tau}

\def\D{\Delta}
\def\F{\Phi}
\def\G{\Gamma}

\def\L{\Lambda}
\def\O{\Omega}

\def\Q{\Theta}

\def\ve{\varepsilon}
\def\vf{\varphi}

\def\cl{{\cal L}}

\def\co{{\cal O}}

\def\bo{{\raise.15ex\hbox{\large$\Box$}}}               
\def\pr{\prod}                                          
\def\ltap{\raisebox{-.4ex}{\rlap{$\sim$}} \raisebox{.4ex}{$<$}}   
\def\gtap{\raisebox{-.4ex}{\rlap{$\sim$}} \raisebox{.4ex}{$>$}}   
\def\face{{\raise.2ex\hbox{$\displaystyle \bigodot$}\mskip-2.2mu \llap {$\ddot
        \smile$}}}                                      
\def\dg{\dagger}                                     
\def\wt#1{\widetilde{#1}}                    
\def\VEV#1{\left\langle #1\right\rangle}        
\def\beq{\begin{equation}}
\def\eeq{\end{equation}}

\def\beqa{\begin{eqnarray}}
\def\eeqa{\end{eqnarray}}
\def\NO{\nonumber}

\def\wt#1{\widetilde{#1}}                    
\def\wh#1{\widehat{#1}}                        
\def\Bar#1{\overline{#1}}                       
\def\VEV#1{\left\langle #1\right\rangle}        
\def\abs#1{\left| #1\right|}                    
\def\leftrightarrowfill{$\mathsurround=0pt \mathord\leftarrow \mkern-6mu
        \cleaders\hbox{$\mkern-2mu \mathord- \mkern-2mu$}\hfill
        \mkern-6mu \mathord\rightarrow$}       
\def\dvec#1{\vbox{\ialign{##\crcr
        \leftrightarrowfill\crcr\noalign{\kern-1pt\nointerlineskip}
        $\hfil\displaystyle{#1}\hfil$\crcr}}}           

\def\pl#1#2#3{Phys.~Lett.~{\bf B {#1}} (19{#2}) #3}
\def\np#1#2#3{Nucl.~Phys.~{\bf B {#1}} (19{#2}) #3}
\def\prl#1#2#3{Phys.~Rev.~Lett.~{\bf #1} (19{#2}) #3}
\def\pr#1#2#3{Phys.~Rev.~{\bf D {#1}} (19{#2}) #3}

\def\nc#1#2#3{Nuovo Cim.~{\bf {#1}} (19{#2}) #3}

\begin{document}
\title{
{\normalsize
\begin{minipage}{5cm}
DESY 99-187\\
UPR-892-T\\
July 2000
\end{minipage}}\hspace{\fill}\mbox{}\\[5ex]
{\bf NEUTRINO MASSES AND\\ THE BARYON ASYMMETRY}
}
\author{W. Buchm\"uller \\
\vspace{3.0\baselineskip}                                               
{\normalsize\it Deutsches Elektronen-Synchrotron DESY, 22603 Hamburg, Germany}
\\
\vspace*{0.1cm}\\
M. Pl\"umacher \\
\vspace{3.0\baselineskip}                                               
{\normalsize\it Department of Physics and Astronomy, University of 
Pennsylvania}\\
{\normalsize\it Philadelphia, PA 19104, U.S.A.}\\
}        
\date{}
\maketitle
\thispagestyle{empty}
\begin{abstract}
\noindent
Due to sphaleron processes in the high-temperature symmetric phase of the
standard model the cosmological baryon asymmetry is related to neutrino
properties. For hierarchical neutrino masses, with $B-L$ broken at the
unification scale $\Lambda_{GUT}\sim 10^{16}\;$GeV,  
the observed baryon asymmetry $n_B/s \sim 10^{-10}$ can be naturally
explained by the decay of heavy Majorana neutrinos. We illustrate this
mechanism with two models of neutrino masses, consistent with the solar
and atmospheric neutrino anomalies, which are based on the two symmetry groups
$SU(5)\times U(1)_F$ and $SU(3)_c\times SU(3)_L\times SU(3)_R\times U(1)_F$.
We also review related cosmological bounds on  Majorana
neutrino masses and the use of Boltzmann equations. 
\end{abstract}

\newpage
\tableofcontents

\newpage

\section{Introduction}

The generation of a cosmological baryon asymmetry can be understood as a
consequence of baryon number violation, C and CP violation, and a deviation 
from thermal equilibrium \cite{sa67}. All these conditions can be naturally
satisfied in the context of unified extensions of the standard model of 
strong and electroweak interactions. In particular the deviation from
thermal equilibrium is realized in the out-of-equilibrium decay of heavy
particles whose mass is related to the mass scale of unification \cite{yo78}. 
The presently observed matter-antimatter asymmetry, the ratio of the baryon 
density to the entropy density of the universe,
\beq
Y_B = {(n_B-n_{\overline{B}}) \over s} = (0.6 - 1)\cdot 10^{-10}\;,
\eeq
is then explained as a consequence of the spectrum and interactions of 
elementary particles, together with the cosmological evolution. A general
overview of different scenarios for baryogenesis can be found in \cite{dol92}.

\begin{figure}[h]
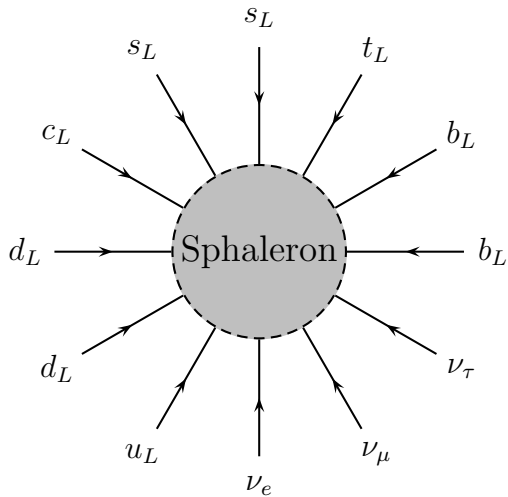

\begin{center}
\scaleboxto(7,7) {\parbox[c]{9cm}{ \begin{center}
     \pspicture*(-0.50,-2.5)(8.5,6.5)
     \psset{linecolor=lightgray}
     \qdisk(4,2){1.5cm}
     \psset{linecolor=black}
     \pscircle[linewidth=1pt,linestyle=dashed](4,2){1.5cm}
     \rput[cc]{0}(4,2){\scalebox{1.5}{Sphaleron}}
     \psline[linewidth=1pt](5.50,2.00)(7.50,2.00)
     \psline[linewidth=1pt](5.30,2.75)(7.03,3.75)
     \psline[linewidth=1pt](4.75,3.30)(5.75,5.03)
     \psline[linewidth=1pt](4.00,3.50)(4.00,5.50)
     \psline[linewidth=1pt](3.25,3.30)(2.25,5.03)
     \psline[linewidth=1pt](2.70,2.75)(0.97,3.75)
     \psline[linewidth=1pt](2.50,2.00)(0.50,2.00)
     \psline[linewidth=1pt](2.70,1.25)(0.97,0.25)
     \psline[linewidth=1pt](3.25,0.70)(2.25,-1.03)
     \psline[linewidth=1pt](4.00,0.50)(4.00,-1.50)
     \psline[linewidth=1pt](4.75,0.70)(5.75,-1.03)
     \psline[linewidth=1pt](5.30,1.25)(7.03,0.25)
     \psline[linewidth=1pt]{<-}(6.50,2.00)(6.60,2.00)
     \psline[linewidth=1pt]{<-}(6.17,3.25)(6.25,3.30)
     \psline[linewidth=1pt]{<-}(5.25,4.17)(5.30,4.25)
     \psline[linewidth=1pt]{<-}(4.00,4.50)(4.00,4.60)
     \psline[linewidth=1pt]{<-}(2.75,4.17)(2.70,4.25)
     \psline[linewidth=1pt]{<-}(1.83,3.25)(1.75,3.30)
     \psline[linewidth=1pt]{<-}(1.50,2.00)(1.40,2.00)
     \psline[linewidth=1pt]{<-}(1.83,0.75)(1.75,0.70)
     \psline[linewidth=1pt]{<-}(2.75,-0.17)(2.70,-0.25)
     \psline[linewidth=1pt]{<-}(4.00,-0.50)(4.00,-0.60)
     \psline[linewidth=1pt]{<-}(5.25,-0.17)(5.30,-0.25)
     \psline[linewidth=1pt]{<-}(6.17,0.75)(6.25,0.70)
     \rput[cc]{0}(8.00,2.00){\scalebox{1.3}{$b_L$}}
     \rput[cc]{0}(7.46,4.00){\scalebox{1.3}{$b_L$}}
     \rput[cc]{0}(6.00,5.46){\scalebox{1.3}{$t_L$}}
     \rput[cc]{0}(4.00,6.00){\scalebox{1.3}{$s_L$}}
     \rput[cc]{0}(2.00,5.46){\scalebox{1.3}{$s_L$}}
     \rput[cc]{0}(0.54,4.00){\scalebox{1.3}{$c_L$}}
     \rput[cc]{0}(0.00,2.00){\scalebox{1.3}{$d_L$}}
     \rput[cc]{0}(0.54,0.00){\scalebox{1.3}{$d_L$}}
     \rput[cc]{0}(2.00,-1.46){\scalebox{1.3}{$u_L$}}
     \rput[cc]{0}(4.00,-2.00){\scalebox{1.3}{$\nu_e$}}
     \rput[cc]{0}(6.00,-1.46){\scalebox{1.3}{$\nu_{\mu}$}}
     \rput[cc]{0}(7.46,0.00){\scalebox{1.3}{$\nu_{\tau}$}}
     \endpspicture
\end{center}}}
\end{center}
\caption{\it One of the 12-fermion processes which are in thermal 
equilibrium in the high-temperature phase of the standard model.
\label{fig_sphal}}
\end{figure}

A crucial ingredient of baryogenesis is the connection between baryon number
($B$) and lepton number ($L$) in the high-temperature, symmetric phase of
the standard model. Due to the chiral nature of the weak interactions $B$ and
$L$ are not conserved \cite{tho76}. At zero temperature this has no observable 
effect due to the smallness of the weak coupling. However, as the temperature 
approaches the critical temperature $T_c$ of the electroweak phase 
transition \cite{ki72}, $B$ and $L$ violating processes come into thermal 
equilibrium \cite{krs85}. 

The rate of these processes is
related to the free energy of sphaleron-type field configurations which carry
topological charge. In the standard model they lead to an effective
interaction of all left-handed fermions \cite{tho76} 
(cf. fig.~\ref{fig_sphal}), 
\beq\label{obl}
O_{B+L} = \prod_i \left(q_{Li} q_{Li} q_{Li} l_{Li}\right)\; ,
\eeq
which violates baryon and lepton number by three units, 
\beq 
    \D B = \D L = 3\;. \label{sphal1}
\eeq
The sphaleron transition rate in the symmetric high-temperature phase
has been evaluated by combining an analytical resummation with numerical
lattice techniques \cite{bmr00}. The result is, in accord with previous 
estimates, that $B$ and $L$ violating processes are in thermal equilibrium for 
temperatures in the range
\beq 
T_{EW} \sim 100\ \mbox{GeV} < T < T_{SPH} \sim 10^{12}\ \mbox{GeV}\;.
\eeq

Sphaleron processes have a profound effect on the generation of the
cosmological baryon asymmetry.  Eq.~\ref{sphal1} suggests that any
$B+L$ asymmetry generated before the electroweak phase transition,
i.e., at temperatures $T>T_{EW}$, will be washed out. However, since
only left-handed fields couple to sphalerons, a non-zero value of
$B+L$ can persist in the high-temperature, symmetric phase if there
exists a non-vanishing $B-L$ asymmetry. An analysis of the chemical potentials
of all particle species in the high-temperature phase yields the following
relation between the baryon asymmetry $Y_B$ and the corresponding
$L$ and $B-L$ asymmetries $Y_L$ and $Y_{B-L}$, respectively \cite{ht90},
\beq\label{basic}
Y_B\ =\ C\ Y_{B-L}\ =\ {C\over C-1}\ Y_L\;.
\eeq
Here $C$ is a number ${\cal O}(1)$. In the standard model with three 
generations and one Higgs doublet one has $C=28/79$. 

\begin{figure}[h]
\begin{center}
\epsfig{file=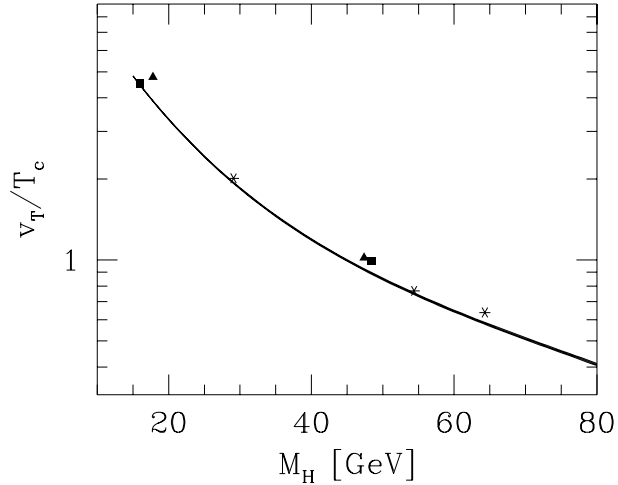,bbllx=63,bblly=280,bburx=560,bbury=670,width=8.2cm}
\end{center}
\caption{\it Jump of the Higgs vacuum expectation value at the critical 
temperature. Comparison of 4d simulations (triangles, squares) 
with 3d simulations (stars) and perturbation theory. \label{jan_fig}}
\end{figure}

An important ingredient in the theory of baryogenesis is also the nature
of the electroweak transition from the high-temperature symmetric phase to the
low-temperature Higgs phase. A first-order phase transition yields a departure
from thermal equilibrium. Since in the standard model baryon number, C and
CP are not conserved, it is conceivable that the cosmological
baryon asymmetry has been generated at the electroweak phase 
transition \cite{krs85}. This possibility has stimulated a large theoretical
activity during the past years to determine the phase diagram of the
electroweak theory. 

Electroweak baryogenesis requires that the baryon asymmetry, which is 
generated during the phase transition, is not erased by sphaleron processes
afterwards. This leads to a condition on the jump of the Higgs vacuum
expectation value $v =\sqrt{\phi^{\dagger}\phi}$ at the critical 
temperature \cite{sh86},
\beq
{\Delta v(T_c)\over T_c} > 1\;.
\eeq
The strength of the electroweak transition has been studied by numerical and
analytical methods as function of the Higgs boson mass. The result for the
$SU(2)$ gauge-Higgs model is shown in fig.~\ref{jan_fig} \cite{ja96}, where
the results of 4d lattice simulations \cite{fhj95}, 3d lattice simulations 
\cite{klr96} and perturbation theory \cite{bfh95} are compared. For Higgs 
masses above 50 GeV this is a good approximation for the full standard model.
Since the present lower bound from LEP on the Higgs mass has reached almost
110 GeV, it is obvious that the electroweak transition in the standard model
is too weak for baryogenesis. For supersymmetric extensions of the standard 
model a sufficiently strong first-order phase transition can still
be achieved for special choices of parameters \cite{cl98}. 

\begin{figure}[h]
\begin{center}
\epsfig{file=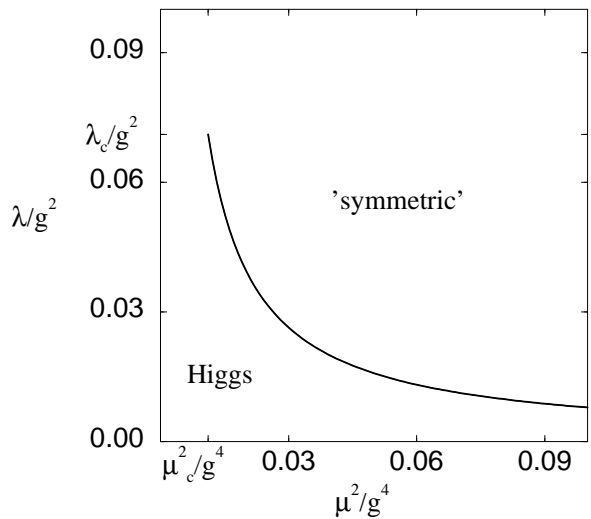,bbllx=48,bblly=262,bburx=510,bbury=687,width=7.5cm}
\end{center}
\caption{\it Critical line of first-order phase transitions.
\label{fig_cross}}
\end{figure}

For large Higgs masses the nature of the electroweak transition is dominated 
by non-perturbative effects of the $SU(2)$ gauge theory at high temperatures. 
At a critical Higgs mass $m_H^c = {\cal O}(m_W)$ an intriguing phenomenon 
occurs: the first-order phase transition turns into a smooth 
crossover \cite{bp95,klrs96,bw97}, as expected on general grounds \cite{ja96}. 
This is shown in fig.~\ref{fig_cross} \cite{bp97} where 
$\lambda/g^2 \simeq m_H^2/(8 m_W^2)$ is plotted as function of 
$\mu^2/ g^4 \simeq (3 m_W^2 + m_H^2)/(64\sqrt{2} G_F m_W^4)(T^2-T_0^2)/T^2$,
with $T_0^2=\sqrt{2}m_H^2/(G_F(3m_W^2+m_H^2))$. For small Higgs masses
one has a first-order phase transition at a critical temperature $T_c(m_H)$.
At the endpoint of the line of first-order transitions, which is reached
for $m_H=m_H^c$, the phase transition is of second order \cite{rtk98}.

The value of the critical Higgs mass can be estimated by comparing the
W-boson mass $m_W$ in the Higgs phase with the magnetic mass $m_{SM}$ in the 
symmetric phase. For $m_{SM}=Cg^2 T$ one obtains \cite{bp97},
\beq\label{cross}
m_H^c = \left({3\over 4\pi C}\right)^{1/2} m_W \simeq 74\ \mbox{GeV}\;,
\eeq
where we have used $C\simeq 0.35$ \cite{eb99}. Numerical lattice simulations
have determined the precise value $m_H^c = 72.1 \pm 1.4$~GeV \cite{fo99}. 
The estimate (\ref{cross}) can also be extended to the supersymmetric
extensions of the standard model, where one obtains for the critical Higgs
mass $m_h^c < 130\dots 150$~GeV \cite{ce97}.

The detailed studies of the electroweak phase transition have shown, that 
for Higgs masses above the present LEP bound of about 110 GeV the
cosmological baryon asymmetry did not change during this transition, except
possibly for a small parameter range in the supersymmetric standard model.
In particular, the electroweak transition may have been just a smooth 
crossover, without any deviation from thermal equilibrium. In this case it's 
sole effect has been to switch off the $B-L$ changing sphaleron processes 
adiabatically. 

Based on the relation (\ref{basic}) between baryon and lepton number we then 
conclude that $B-L$ violation is needed to explain the cosmological baryon
asymmetry if baryogenesis took place before the electroweak transition, i.e. 
at temperatures $T > T_{EW} \sim 100$~GeV. In the standard model, as well as 
its supersymmetric version and its unified extensions based on the gauge group 
$SU(5)$, $B-L$ is a conserved quantity. Hence, no baryon asymmetry can be 
generated dynamically in these models and one has to consider extensions
with lepton number violation.

\begin{figure}[h]
\begin{center}
\scaleboxto(7,3.5){
\parbox[c]{9cm}{\input{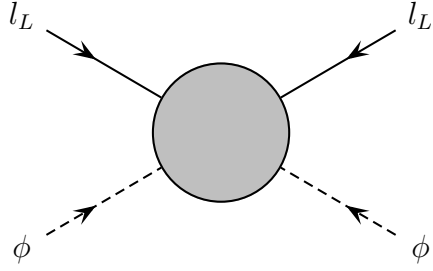}}}
\end{center}
\caption{\it Effective lepton number violating interaction.\label{fig_lept}}
\end{figure}

The remnant of lepton number violation at low energies is the appearance of an
effective $\Delta L=2$ interaction between lepton and Higgs fields
(cf.~fig.~\ref{fig_lept}),
\beq\label{dl2} 
\cl_{\Delta L=2} ={1\over 2} f_{ij}\ l^T_{Li}\vf\ C\ l_{Lj}\vf 
                  +\mbox{ h.c.}\;.\label{intl2}
\eeq
Such an interaction arises in particular from the 
exchange of heavy Majorana neutrinos. In the Higgs phase of the standard
model, where the Higgs field acquires a vacuum expectation value, it gives
rise to Majorana masses of the light neutrinos $\n_e$, $\n_\m$ and $\n_\t$.   

Lepton number violation appears to be necessary to understand the cosmological
baryon asymmetry. However, as we shall see in the following section, the
required lepton number violation can only be weak, otherwise any baryon 
asymmetry would be washed out. These two conditions are the basis of the
interesting contraints which the existence of the matter-antimatter asymmetry
imposes on neutrino physics and on extensions of the standard model in 
general.

\section{Cosmological bounds on neutrino masses}

In the standard model (SM) neutrinos are massless. However, the exchange of
heavy particles can give rise to an effective lepton Higgs interaction
which, after electroweak symmetry breaking, generates neutrino masses.
The lagrangian describing all fermion-Higgs couplings then reads 
\beq\label{inter}
\cl_Y = - h_{d ij}^T\Bar{d_R}_i q_{Lj} H_1
        - h_{u ij}^T\Bar{u_R}_i q_{Lj} H_2
        - h_{e ij}\Bar{e_R}_i l_{Lj} H_1
        + {1\over 2} f_{ij}\ l^T_{Li}H_2\ C\ l_{Lj}H_2 +\mbox{ h.c.}\;.
\eeq
Here $q_{Li}$, $u_{Ri}$, $d_{Ri}$, $l_{Li}$, $e_{Ri}$, $i=1\ldots N$, are N 
generations of quark and lepton fields, $H_1$ and $H_2$ are Higgs fields with 
vacuum expectation values $v_i=\VEV{H_i^0}\ne0$. $h_d$, $h_u$, $h_e$ and $f$ 
are N$\times$N complex matrices. For the further discussion it is convenient 
to choose a basis where $h_u$ and $h_e$ are diagonal and real. In the case
$v_1 \sim v_2 \sim v=\sqrt{v_1^2+v_2^2}$ the smallest Yukawa coupling is 
$h_{e11}=m_e/v_1$ followed by $h_{u11}=m_u/v_2$. 

The mixing of the different quark generations is given by the 
Kobayashi-Maskawa matrix $V_d$, which is defined by
\beq
 h_d^T\ V_d\ = \ {m_d\over v_1}\;.
\eeq
Here $m_d$ is the diagonal real down quark mass matrix, and the weak 
eigenstates of the right-handed d-quarks have been chosen to be identical to
the mass eigenstates. Correspondingly,
the mixing matrix $V_\n$ in the leptonic charged current is determined by
\beq
 V_\n^T\ f\ V_\n\ =\ -\ {m_\n \over v_2^2}\;, 
 \label{vnu}
\eeq
where 
\beq
m_\n=\left(\begin{array}{ccc}m_1&0&0\\0&m_2&0\\0&0&m_3
              \end{array}\right)
\eeq 
is the diagonal and real mass matrix of the light Majorana neutrinos.

\subsection{Chemical equilibrium}
  
In a weakly coupled plasma with temperature $T$ and volume $V$ one can
assign a chemical potential $\m$ to each of the quark, lepton and Higgs
fields. In the SM with one Higgs doublet, i.e., $H_2=\wt H_1 \equiv \vf$,
and N generations one has 5N+1 chemical potentials. The corresponding 
partition function is \cite{lan}
\beq
Z(\m,T,V) = \mbox{Tr}e^{-\b(H-\sum_i\m_i Q_i)}\;.
\eeq
Here $\b=1/T$, $H$ is the Hamilton operator and $Q_i$ are the charge 
operators for the corresponding fields. The asymmetry in the particle and
antiparticle number densities is then given by the derivative of the 
thermodynamic potential,
\beq
n_i-\overline{n}_i=-{\partial \O(\m,T)\over \partial \m_i}\;, \quad
\O(\m,T) = - {T\over V} \ln{Z(\m,T,V)}\;.
\eeq
For a non-interacting gas of massless particles one has
\beq\label{number}
n_i-\overline{n}_i={g T^3\over 6}
\left\{\begin{array}{rl}\b\m_i +{\cal O}\left(\left(\b\m_i\right)^3\right)\;,
&\mbox{fermions}\;,\\
2\b\m_i+{\cal O}\left(\left(\b\m_i\right)^3\right)\;, &\mbox{bosons}\;.
\end{array}\right.
\eeq
The following analysis will be based on these relations for $\b \m_i \ll 1$.
However, one should keep in mind that the plasma of the early universe is
very different from a weakly coupled relativistic gas due to the presence
of unscreened non-abelian gauge interactions. Hence, non-perturbative effects
may be important in some cases.

In the high-temperature plasma quarks, leptons and Higgs bosons interact via
Yukawa and gauge couplings and, in addition, via the non-perturbative
sphaleron processes. In thermal equilibrium all these processes yield
constraints between the various chemical potentials. The effective interaction
(\ref{obl}) induced by the $SU(2)$ electroweak instantons yields the 
constraint \cite{krs85},
\beq\label{sphew}
\sum_i\left(3\m_{qi} + \m_{li}\right) = 0\;.
\eeq
One also has to take the $SU(3)$ QCD instanton processes into account 
\cite{moh} which generate the effective interaction
\beq\label{oaxi}
O_A = \prod_i \left(q_{Li}q_{Li}u^c_{Ri}d^c_{Ri}\right)
\eeq
between left-handed and right-handed quarks. The corresponding relation
between the chemical potentials reads
\beq\label{sphqcd}
\sum_i\left(2\m_{qi} - \m_{ui} - \m_{di}\right) = 0\;.
\eeq
A third condition, which is valid at all temperatures, arises from the
requirement that the total hypercharge of the plasma vanishes. From
eq.~(\ref{number}) and the known hypercharges one obtains
\beq\label{hypsm}
\sum_i\left(\m_{qi} + 2 \m_{ui} - \m_{di} - \m_{li} - \m_{ei} + 
{2\over N} \m_{\vf}\right) = 0\;.
\eeq

The Yukawa interactions, supplemented by gauge interactions, yield relations
between the chemical potentials of left-handed and right-handed fermions,
\beq\label{myuk}
\m_{qi}-\m_{\vf}-\m_{dj} = 0\;, \quad
\m_{qi}+\m_{\vf}-\m_{uj} = 0\;, \quad
\m_{li}-\m_{\vf}-\m_{ej} = 0\;.
\eeq
Furthermore, the $\D L =2$ interaction in (\ref{inter}) implies
\beq\label{mdl2}
\m_{li}+\m_{\vf} = 0\;.
\eeq

The above relations between chemical potentials hold if the corresponding
interactions are in thermal equilibrium. In the temperature range
$T_{EW} \sim 100\ \mbox{GeV} < T < T_{SPH} \sim 10^{12}\ \mbox{GeV}$, 
which is of interest for 
baryogenesis, this is the case for all gauge interactions. It is not always 
true, however, for Yukawa interactions. The rate of a scattering process 
between left- and right-handed fermions, Higgs boson and W-boson, 
\beq
\psi_L \vf \rightarrow \psi_R W\;,
\eeq
is $\G \sim \a \l^2 T$, with $\a=g^2/(4\pi)$. This rate has to be
compared with the Hubble rate,
\beq
H \simeq 0.33\ g_*^{1/2} {T^2\over M_{PL}} 
  \simeq 0.1\ g_*^{1/2} {T^2\over 10^{18}\ \mbox{GeV}}\;.
\eeq  
The equilibrium condition $\G(T) > H(T)$ is satisfied for sufficiently small 
temperatures, 
\beq\label{eqyuk}
T < T_\l \sim \l^2 10^{16}\ \mbox{GeV}\; .
\eeq
Hence, one obtains for the decoupling temperatures of right-handed electrons,
up-quarks,...,
\beq
T_e \sim 10^4\ \mbox{GeV}\;,\quad T_u \sim 10^6\ \mbox{GeV}\;,\ldots .
\eeq
At a temperature $T \sim 10^{10}$ GeV, which is characteristic of leptogenesis,
$e_R\equiv e_{R1}$, $\m_R\equiv e_{R2}$, $d_R\equiv d_{R1}$, $s_R\equiv d_{R2}$
and $u_R\equiv u_{R1}$ are out of equilibrium.

Note that the QCD sphaleron constraint (\ref{sphqcd}) is automatically
satisfied if the quark Yukawa interactions are in equilibrium 
(cf.~(\ref{myuk})). If the Yukawa interaction of one of the right-handed 
quarks is too weak, the sphaleron constraint still establishes full
chemical equilibrium. 

Using eq.~(\ref{number}) also the baryon number density $n_B \equiv B T^2/6$
and the lepton number densities $n_L \equiv L T^2/6$ can be expressed in terms
of the chemical potentials. The baryon asymmetry $B$ and the lepton asymmetries
$L_i$ read
\beqa
B &=& \sum_i \left(2\m_{qi} + \m_{ui} + \m_{di}\right)\;, \\
L_i &=& 2\m_{li} + \m_{ei}\;,\quad L=\sum_i L_i\;.
\eeqa

\subsection{Relations between $B$, $L$ and $B-L$}

Knowing which particle species are in thermal equilibrium one can derive
relations between different asymmetries. Consider first the most familiar
case where all Yukawa interactions are in equilibrium and the $\D L=2$
lepton-Higgs interaction is out of equilibrium. In this case the asymmetries 
$L_i-B/N$ are conserved. The Yukawa interactions establish equilibrium 
between the different generations,
\beq
\m_{li} \equiv \m_l\;, \quad \m_{qi} \equiv \m_q\;, \quad \mbox{etc.}
\eeq  
Together with the sphaleron process and the hypercharge constraint they
allow to express all chemical potentials, and therefore all asymmetries,
in terms of a single chemical potential which may be chosen to be $\m_l$.
The result reads
\beqa\label{exam1}
\m_e &=& {2N+3\over 6N+3}\m_l\;, \quad
\m_d = -{6N+1\over 6N+3}\m_l\;, \quad
\m_u = {2N-1\over 6N+3}\m_l\;, \NO\\
\m_q &=& -{1\over 3} \m_l\;, \quad
\m_{\vf} = {4N\over 6N+3} \m_l\;.
\eeqa
The corresponding baryon and lepton asymmetries are
\beq
B = -{4N\over 3}\m_l\;, \quad L = {14N^2+9N\over 6N+3}\m_l\;,
\eeq
which yields the well-known connection between the $B$ and $B-L$ 
asymmetries \cite{ks}
\beq
B = {8N+4\over 22N+13} (B-L)\;.
\eeq
Note, that this relation only holds for temperatures $T\gg v$. In general,
the ratio $B/(B-L)$ is a function of $v/T$ \cite{ks1,ls00}.

Another instructive example is the case where the $\D L=2$ interactions are
in equilibrium but the right-handed electrons are not. Depending on the 
neutrino masses and mixings, this could be the case for temperatures above
$T_e \sim 10^4$ GeV \cite{cko93}. Right-handed electron number would then be
conserved, and Yukawa and gauge interactions would relate all asymmetries
to the asymmetry of right-handed electrons. The various chemical potentials
are given by ($\m_e = \m_{e1}$, $\m_{\bar{e}} = \m_{e2}=\ldots=\m_{eN}$),
\beqa\label{exam2}
\m_{\bar{e}}&=&-{3\over 10 N}\m_e\;,\quad 
\m_d = -{1\over 10 N}\m_e\;,\quad
\m_u = {1\over 5 N}\m_e\;, \NO\\
\m_l&=&-{3\over 20 N}\m_e\;,\quad
\m_q = {1\over 20 N}\m_e\;, \quad
\m_{\vf} = {3\over 20 N}\m_e\;.
\eeqa
The corresponding baryon and lepton asymmetries are \cite{cko93}
\beq
B = {1\over 5}\m_e \;, \quad L = {4N+3\over 10 N}\m_e\;,
\eeq
which yields for the relation between $B$ and $B-L$,
\beq
B = -{2N\over 2N+3} (B-L)\;.
\eeq
Note that although sphaleron processes and $\D L=2$ processes are in
equilibrium, the asymmetries in $B$, $L$ and $B-L$ do not vanish! 

\subsection{Constraint on Majorana neutrino masses}

The two examples illustrate the connection between lepton number and
baryon number induced by sphaleron processes. They also show how this
connection depends on other processes in the high-temperature plasma.
To have one quark-Higgs or lepton-Higgs interaction out of equilibrium
is sufficient in order to have non-vanishing $B$, $L$ and $B-L$. If
all interactions in (\ref{inter}) are in equilibrium, eqs.~(\ref{mdl2})
and (\ref{exam1}) together imply $\m_l=0$ and therefore
\beq
B = L = B-L = 0\;,
\eeq
which is inconsistent with the existence of a matter-antimatter asymmetry.
Since the equilibrium conditions of the various interactions are
temperature dependent, and the $\D L=2$ interaction is related to neutrino
masses and mixings, one obtains important constraints on neutrino properties
from the existence of the cosmological baryon asymmetry.

The $\D L=2$ processes described by (\ref{dl2}) take place with the 
rate \cite{fy1}
  \beq
    \Gamma_{\Delta L=2} (T) = {1\over \pi^3}\,{T^3\over v^4}\, 
    \sum_{i=e,\m,\t} m_{\n_i}^2\; .
  \eeq
Requiring $\G_{\D L=2}(T) < H(T)$ then yields an upper bound on Majorana 
neutrino masses,
\beq\label{nbound}
\sum_i m_{\n_i}^2 < \left(0.2\ \mbox{eV}\ \left({T_{SPH}\over T}
                    \right)^{1/2} \right)^2\;.
\eeq
For typical leptogenesis temperatures $T \sim 10^{10}$~GeV this bound is 
comparable to the upper bound on the electron neutrino mass 
obtained from neutrinoless double beta decay. Note, that the bound
also applies to the $\t$-neutrino mass. However, if one uses for $T$ the
decoupling temperature of right-handed electrons, $T_e \sim 10^4$~GeV, the
much weaker bound $m_\n < 2$~keV is obtained \cite{cko93}.

Clearly, what temperature one has to use in eq.~(\ref{nbound}) depends on
the thermal history of the early universe. Some information is needed on
what kind of asymmetries may have been generated as the temperature decreased.
This, together with the temperature dependence of the lepton-Higgs interactions
then yields constraints on neutrino masses. 

\subsection{Primordial asymmetries}

The possible generation of asymmetries can be systematically studied by
listing all the higher-dimensional $SU(3)\times SU(2)\times U(1)$
operators which may be generated by
the exchange of heavy particles. The dynamics of the heavy particles may
then generate an asymmetry in the quantum numbers carried by the massless
fields which appear in the operator.

For d=5, there is a unique operator, which has already been discussed
above,
\beq
(l_{Li}\vf)(l_{Lj}\vf)\;.
\eeq
It is generated in particular by the exchange of heavy Majorana neutrinos
whose coupling to the massless fields is
\beq
h_{\n ij}\Bar{\n_R}_i l_{Lj} \vf \;.
\eeq
The out-of-equilibrium decays of the heavy neutrinos can generate a lepton 
asymmetry, which is the well-known mechanism of leptogenesis. The decays
yield asymmetries $L_i-B/N$ which are conserved in the subsequent evolution.  
The initial asymmetry in right-handed electrons is zero. In order to satisfy 
the out-of-equilibrium condition it is very important that
at least some Yukawa couplings are small and that the right-handed neutrinos
carry no quantum numbers with respect to unbroken gauge symmetries.

In order to study possible asymmetries of right-handed electrons one has
to consider operators containing $e_R$. A simple example, with d=6, reads
\beq
(q_{Li}l_{Lj})(u^c_{Rk}e^c_{Rl})\;.
\eeq
It can be generated by leptoquark exchange ($\chi \sim (3^*,1,1/3)$),
\beq
(\l^q_{ij} q^T_{Li}C l_{Lj} + \l^u_{ij} u^T_{Ri}C e_{Rj})\chi\;.
\eeq
Note, that $\wt{\vf}$ and $\chi$ form a $5^*$-plet of $SU(5)$. In principle,
out-of-equilibrium decays of leptoquarks may generate a $e_R$ asymmetry.
One may worry, however, whether the branching ratio into final states
containing $e_R$ is sufficiently large. Furthermore, it appears very
difficult to satisfy the out-of-equilibrium condition since leptoquarks
carry colour. Maybe, all these problems can be overcome by making use of
coherent oscillations of scalar fields \cite{ad85} or by special particle 
production mechanisms after inflation. However, we are not aware of a
consistent scenario for the generation of a $e_R$-asymmetry. 
Hence, it appears appropriate to take the bound
eq.~(\ref{nbound}) on Majorana neutrino masses as a guideline and to examine
its validity in each particular model.

\subsection{Supersymmetry}

At temperatures $T$ where some of the interactions are out of thermal
equilibrium the effective theory acquires a larger symmetry. Examples
discussed above are the $\D L =2$ interactions and the Yukawa couplings
of the right-handed electron. In the latter case the symmetry of the
effective theory is generated by $Q_{eR}$, the number operator
of right-handed electrons. As a consequence, the chemical potential $\m_e$,
i.e. the $e_R$-asymmetry, is a free parameter whose value influences $B$
and $L$.

In the supersymmetric standard model (SSM) one has Yukawa interactions given by
the superpotential
\beq\label{superp}
W  = h_{dij}D^c_i Q_j H_1 + h_{uij}U^c_i Q_j H_2
     + h_{eij}E^c_i L_j H_1 + {1\over 2} f_{ij} L_i H_2 L_j H_2 \;,
\eeq
and mass parameters, the supersymmetric mass term $\m H_1 H_2$ and soft 
supersymmetry breaking scalar masses and gaugino masses,
\beq
L_m = {1\over 2}m_g \tilde{g}_L\tilde{g}_L
      +{1\over 2}m_w \tilde{w}_L\tilde{w}_L
      +{1\over 2}m_b \tilde{b}_L\tilde{b}_L\ +\ \mbox{h.c.}\;,
\eeq
for bino, wino and gluino, respectively. Naturally, the gaugino masses
and the $\m$-parameter are of order the gravitino mass,
$m_i \sim \m \sim m_{\tilde{G}}$. For unbroken supersymmetry, the 
SSM lagrangian has two chiral U(1) symmetries in addition to those of the SM,
an axial Peccei-Quinn symmetry and R-invariance.

Due to the additional fermions, the higgsinos $h_{L1},h_{L2}$ and the 
gauginos, the effective instanton induced interactions are modified. For
a Weyl fermion $\psi$ in the adjoint representation of a $SU(N)$ gauge theory 
the 't Hooft determinant reads,
\beq
O_{adj} =\psi_{L_1}\ldots \psi_{L_{2N}}\;.
\eeq
For a theory with Weyl fermions in the fundamental and the adjoint
representation the `t Hooft determinant is the product of the two single
contributions. Hence, in the SSM eqs.~(\ref{obl}) and (\ref{oaxi}) are 
replaced by \cite{iq}
\beq
\wh O_{B+L} = \prod_i \left(q_{Li}q_{Li}q_{Li}l_{Li}\right)
      \tilde{h}_{L1}\tilde{h}_{L2}
      \tilde{w}_L\tilde{w}_L\tilde{w}_L\tilde{w}_L
\eeq
and
\beq
\wh O_A = \prod_i \left(q_{Li}q_{Li}u^c_{Ri}d^c_{Ri}\right)
      \tilde{g}_L\tilde{g}_L\tilde{g}_L\tilde{g}_L\tilde{g}_L\tilde{g}_L\;,
\eeq
respectively. The corresponding relations for the chemical potentials read,
\beq\label{sph2}
\sum_i\left(3\m_{qi}+\m_{li}\right)+\m_{\tilde{h}1}+\m_{\tilde{h}2}
+4\m_{\tilde{w}} = 0\;,
\eeq
and
\beq\label{sph3}
\sum_i\left(2\m_{qi} - \m_{ui} - \m_{di}\right) + 6\m_{\tilde{g}} = 0\;.
\eeq
Due to the different Higgs content of the SSM also the zero hypercharge
constraint is modified,
\beqa\label{hypssm}
0 &=& \sum_i\left(\m_{qi} + 2 \m_{ui} - \m_{di} - \m_{li} - \m_{ei}\right) + 
{1\over N}(\m_{h2}- \m_{h1})  \NO\\
&& +2\sum_i\left(\m_{\tilde{Qi}} + 2 \m_{\tilde{ui}} - \m_{\tilde{di}} - 
\m_{\tilde{li}} - \m_{\tilde{ei}}\right) + 
{1\over 2N}(\m_{\tilde{h}2}- \m_{\tilde{h}1}) \;.
\eeqa
Note, that scalars contribute twice as much as fermions.
Furthermore, one obtains from the gaugino interactions,
\beq
\m_{\tilde{Q}i}=\m_{\tilde{b}}+\m_{qi}=\m_{\tilde{w}}+\m_{qi}
               =\m_{\tilde{g}}+\m_{qi}\;,
\eeq
and therefore,
\beq
\m_{\tilde{b}}=\m_{\tilde{w}}=\m_{\tilde{g}}\;.
\eeq
The gaugino interactions also determine all chemical potentials of scalars
in terms of the chemical potentials of the corresponding fermions.

If all mass terms, i.e., all effects of supersymmetry breaking, are in 
equilibrium, one has
\beq
\m_{\tilde{h}1}+\m_{\tilde{h}2}=0\;,
\quad \m_{\tilde{b}}=\m_{\tilde{w}}=\m_{\tilde{g}}=0\;.
\eeq
The number of parameters is then the same as in the SM and one obtains
essentially the same relations between $B$, $L$ and $B-L$.

However, at sufficiently high temperatures, $T>T_{SB}$, supersymmetry
breaking effects will not be in equilibrium. As an example, consider the
case that all Yukawa interactions are in equilibrium, which implies
$\m_{li}=\m_l$, $\m_{qi}=\m_q$, etc. The above relations then determine
all chemical potentials in terms of $\m_l$ and $\m_{\tilde{g}}$ \cite{iq}. 
The baryon and lepton asymmetries
\beqa
B &=&  N \left(2\m_{q} + \m_{u} + \m_{d}\right)
+ 2 N \left(2\m_{\tilde{Q}} + \m_{\tilde{u}} + \m_{\tilde{d}}\right)\;, \\
L &=& N\left(2\m_{L} + \m_{e}\right)
+ 2 N \left( 2\m_{\tilde{L}} + \m_{\tilde{e}}\right)\;,
\eeqa
are then given by,
\beqa
B &=& -4N\m_l + {10N-24\over N}\m_{\tilde{g}}\;,\\
L &=& {14N^2+9N\over 2N+1}\m_l + {4N^2+18N+3\over 2N+1}\m_{\tilde{g}}\;.
\eeqa
Clearly, the simple proportionality between $B$, $L$ and $B-L$ is lost.
In particular, one obtains for $B-L=0$,
\beq
B = {16N^3 + 212 N^2 - 234 N - 216\over 22 N^2 + 13 N} \m_{\tilde{g}}\;,
\eeq
which yields $B=6\m_{\tilde{g}}$ for $N=3$.
Hence, at temperatures $T>T_{SB}$ also an asymmetry in gauginos affects the
baryon asymmetry. 

\subsection{R-parity violating interactions}

Like for Majorana neutrinos, cosmological bounds can also be derived on 
the strength of new baryon and lepton number changing interactions \cite{cdx,
fgx91,dr}. Of particular interest are R-parity violating interactions in 
supersymmetric theories which allow single production of supersymmetric 
particles at colliders.

Consider, as an example, the $\D L=1$ Yukawa couplings
\beq\label{delr}
W_{\D R} = \l_{ijk} D_i^c Q_j L_k\;,
\eeq
which may appear natural since the Higgs multiplet $H_1$ and the lepton
multiplets $L_k$ have the same gauge quantum numbers. Requiring 
$T_\l < T_{EW} \sim 100$ GeV and neglecting the masses of scalar quarks
and leptons, one obtains from eq.~(\ref{eqyuk})
\beq\label{bour}
\l < 10^{-7}\; ,
\eeq
which would make the interaction (\ref{delr}) unobservable in collider
experiments. A detailed study \cite{dr} of $2\rightarrow 1$ and 
$2\rightarrow 2$ processes taking finite mass effects into account yields 
bounds slightly less stringent than (\ref{bour}).   

It is important, however, that the bound (\ref{bour}) does not apply to
all couplings $\l_{ijk}$ \cite{dr}. It is sufficient that $B-3L_i$ is
effectively conserved for a single lepton flavour. If the couplings
$\l_{ijk}$ have such a flavour structure and if a $B-3L_i$ asymmetry is indeed
generated a primordial baryon asymmetry can be preserved. Since the
corresponding couplings $\l$ are smaller than all other Yukawa interactions,
the generation of a $B-3L_i$ asymmetry is indeed conceivable.

\subsection{Finite mass effects}

The relation (\ref{number}) between asymmetries and chemical potentials
holds for an ideal relativistic gas. It is modified due to interactions.
To leading order in the couplings their effect can be expressed in terms
of thermal masses \cite{dr,krs2,dko94}. In the symmetric, 
high-temperature phase one finds \cite{dko94},
\beq\label{number1}
n_i-\overline{n}_i={g T^2\over 6}\m_i \left(1 - {3\over \p^2}
{m_i^2(T)\over T^2} \ldots \right)\; .
\eeq
For leptons, for instance, the thermal masses are given by
\beqa
m_{Li}^2(T) &=& \left({3\over32}g_2^2 + {1\over 32}g_1^2 
               + {1\over 6}h_{ei}^2\right)\;,\nonumber\\
m_{Ri}^2(T) &=& \left({1\over 8}g_1^2 + {1\over 8}h_{ei}^2\right)\;,
\eeqa
where $g_1$ and $g_2$ are the $U(1)$ and $SU(2)$ gauge couplings, respectively.

The thermal masses also affect the relation between baryon and lepton
numbers. In particular, one can have $B\neq 0$ with $B-L=0$ if the lepton
asymmetry is flavour dependent. In the symmetric phase one finds \cite{dko94},
\beqa
B &=& {517\over 2844 \pi^2} h_\t^2 \left(L_e-L_\t +L_\m-L_\t\right)\nonumber\\
&\simeq & 9\cdot 10^{-7} \left(L_e-L_\t +L_\m-L_\t\right)\;.
\eeqa
Although this finite mass effect is very small, it can have interesting
consequences if large flavour dependent lepton asymmetries are generated 
in a theory where $B-L$ is conserved.  

\newpage

\section{Baryogenesis and neutrino masses}

\subsection[Baryogenesis through Majorana neutrino decays]{Baryogenesis through Majorana neutrino decays \protect\label{nudecay}}
  As discussed in section~1, baryogenesis before the electroweak transition 
requires $B-L$ violation, and therefore $L$ violation. Lepton number
  violation is most simply realized by adding right-handed Majorana
  neutrinos to the standard model.  Heavy right-handed Majorana
  neutrinos, whose existence is predicted by all extensions of the
  standard model containing $B-L$ as a local symmetry, can also
  explain the smallness of the light neutrino masses via the see-saw
  mechanism \cite{seesaw}.

  The most general Lagrangian for couplings and masses of charged
  leptons and neutrinos reads
  \beq\label{yuk}
  \cl_Y = -h_{e ij}\Bar{e_R}_i l_{Lj} H_1 
          -h_{\n ij}\Bar{\n_R}_i l_{Lj} H_2
          -{1\over2}h_{r ij} \Bar{\n^c_R}_i \n_{Rj} R +\mbox{ h.c.}\;.
  \eeq
  The vacuum expectation values of the Higgs fields, $\VEV{H_1}=v_1$ and
  $\VEV{H_2}=v_2=\tan{\b}\ v_1$, generate Dirac masses $m_e$ and $m_D$
  for charged leptons and neutrinos, $m_e=h_e v_1$ and
  $m_D=h_{\n}v_2$, respectively, which are assumed to be much smaller
  than the Majorana masses $M = h_r\VEV{R}$.  This yields light and
  heavy neutrino mass eigenstates
  \beq
     \n\simeq V_{\nu}^T\n_L+\n_L^c V_{\nu}^*\quad,\qquad
     N\simeq\n_R+\n_R^c\, ,
  \eeq
  with masses
  \beq
     m_{\n}\simeq- V_{\nu}^Tm_D^T{1\over M}m_D V_{\nu}\,
     \quad,\quad  m_N\simeq M\, .
     \label{seesaw}
  \eeq
  Here $V_{\nu}$ is the mixing matrix in the leptonic charged current
  (cf.\ eqs.~(\ref{inter}) and (\ref{vnu})).
  
  The right-handed neutrinos, whose exchange may erase any lepton
  asymmetry, can also generate a lepton asymmetry by means of
  out-of-equilibrium decays. This lepton asymmetry is then partially 
  transformed into a baryon asymmetry by sphaleron processes \cite{fy86}.  
  The decay width of the heavy neutrino $N_i$ reads at tree level,
  \beq
    \G_{Di}=\G\left(N_i\to H_2+l\right)+\G\left(N_i\to H_2^c+l^c\right)
           ={1\over8\p}(h_\n h_\n^\dg)_{ii} M_i\;.
    \label{decay}
  \eeq
A necessary requirement for baryogenesis is the out-of-equilibrium condition
$\Gamma_{D1}< H|_{T=M_1}$ \cite{kt90}, where $H$ is the Hubble parameter
at temperature $T$. From the decay width (\ref{decay}) one then obtains an 
upper bound on an effective light neutrino mass \cite{fgx91,by93}, 
\beqa
    \wt{m}_1\ &=&\ (h_\n h_\n^\dg)_{11} {v_2^2\over M_1}\ 
    \simeq 4 g_*^{1/2} {v_2^2\over M_P}
    \left.{\G_{D1}\over H}\right|_{T=M_1}\NO\\
    &<& 10^{-3}\, \mbox{eV}\;.\label{ooeb}
\eeqa
Here $g_*$ is the number of relativistic degrees of freedom, 
$M_P=(8\pi G_N)^{-1/2}\simeq 2.4\cdot 10^{18}$~GeV is the Planck mass, and 
we have assumed $g_*\simeq 100$, $v_2\simeq 174$~GeV.
  More direct bounds on the light neutrino masses depend on the structure
  of the Dirac neutrino mass matrix. The bound (\ref{ooeb}) implies that
  the heavy neutrinos are not able to follow the rapid change of the
  equilibrium particle distribution once the temperature of the
  universe drops below the mass $M_1$. Hence, the deviation from
  thermal equilibrium consists in a too large number densities of heavy 
neutrinos, as compared to the equilibrium density. 

  \begin{figure}[t]
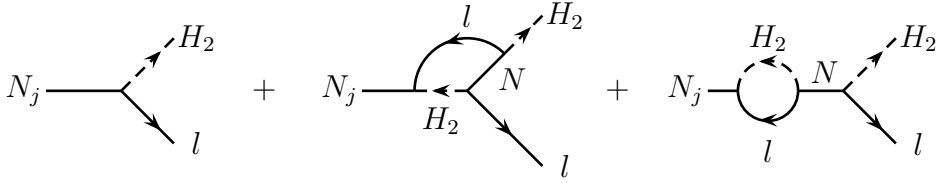

    \centerline{\parbox[c]{12.5cm}{
\pspicture(0,0)(3.7,2.6)
\psline[linewidth=1pt](0.6,1.3)(1.6,1.3)
\psline[linewidth=1pt](1.6,1.3)(2.3,0.6)
\psline[linewidth=1pt,linestyle=dashed](1.6,1.3)(2.3,2.0)
\psline[linewidth=1pt]{->}(2.03,0.87)(2.13,0.77)
\psline[linewidth=1pt]{->}(2.03,1.73)(2.13,1.83)
\rput[cc]{0}(0.3,1.3){$N_j$}
\rput[cc]{0}(2.6,0.6){$l$}
\rput[cc]{0}(2.6,2.0){$H_2$}
\rput[cc]{0}(3.5,1.3){$+$}
\endpspicture
\pspicture(-0.5,0)(4.2,2.6)
\psline[linewidth=1pt](0.6,1.3)(1.3,1.3)
\psline[linewidth=1pt,linestyle=dashed](1.3,1.3)(2.0,1.3)
\psline[linewidth=1pt](2,1.3)(2.5,1.8)
\psline[linewidth=1pt,linestyle=dashed](2.5,1.8)(3,2.3)
\psline[linewidth=1pt](2,1.3)(3,0.3)
\psarc[linewidth=1pt](2,1.3){0.7}{45}{180}
\psline[linewidth=1pt]{<-}(1.53,1.3)(1.63,1.3)
\psline[linewidth=1pt]{<-}(1.7,1.93)(1.8,1.96)
\psline[linewidth=1pt]{->}(2.75,2.05)(2.85,2.15)
\psline[linewidth=1pt]{->}(2.5,0.8)(2.6,0.7)
\rput[cc]{0}(0.3,1.3){$N_j$}
\rput[cc]{0}(1.65,0.9){$H_2$}
\rput[cc]{0}(2,2.3){$l$}
\rput[cc]{0}(2.6,1.45){$N$}
\rput[cc]{0}(3.3,2.3){$H_2$}
\rput[cc]{0}(3.3,0.3){$l$}
\rput[cc]{0}(4.0,1.3){$+$}
\endpspicture
\pspicture(-0.5,0)(3.5,2.6)
\psline[linewidth=1pt](0.5,1.3)(0.9,1.3)
\psline[linewidth=1pt](1.7,1.3)(2.3,1.3)
\psarc[linewidth=1pt](1.3,1.3){0.4}{-180}{0}
\psarc[linewidth=1pt,linestyle=dashed](1.3,1.3){0.4}{0}{180}
\psline[linewidth=1pt]{<-}(1.18,1.69)(1.28,1.69)
\psline[linewidth=1pt]{<-}(1.18,0.91)(1.28,0.91)
\psline[linewidth=1pt](2.3,1.3)(3.0,0.6)
\psline[linewidth=1pt,linestyle=dashed](2.3,1.3)(3.0,2.0)
\psline[linewidth=1pt]{->}(2.73,0.87)(2.83,0.77)
\psline[linewidth=1pt]{->}(2.73,1.73)(2.83,1.83)
\rput[cc]{0}(1.3,0.5){$l$}
\rput[cc]{0}(1.3,2){$H_2$}
\rput[cc]{0}(2.05,1.55){$N$}
\rput[cc]{0}(0.2,1.3){$N_j$}
\rput[cc]{0}(3.3,0.6){$l$}
\rput[cc]{0}(3.3,2.0){$H_2$}
\endpspicture
}}
    \caption{\it Tree level and one-loop diagrams contributing to heavy
    neutrino decays. \label{decay_fig}}
  \end{figure}

  Eventually, however, the neutrinos will decay, and a lepton
  asymmetry is generated due to the CP asymmetry which comes about
  through interference between the tree-level amplitude and the
  one-loop diagrams shown in fig.~\ref{decay_fig}.  In a basis, where
  the right-handed neutrino mass matrix $M = h_r\VEV{R}$ is diagonal,
  one obtains
  \beqa
    \ve_1&=&
    {\Gamma(N_1\rightarrow l \, H_2)-\Gamma(N_1\rightarrow l^c \, H_2^c)
    \over
    \Gamma(N_1\rightarrow l \, H_2)+\Gamma(N_1\rightarrow l^c \, H_2^c)}
    \NO\\[1ex]    
    &\simeq&{1\over8\pi}\;{1\over\left(h_\n h_\n^\dg\right)_{11}}
    \sum_{i=2,3}\mbox{Im}\left[\left(h_\n h_\n^\dg\right)_{1i}^2\right]
    \left[f\left(M_i^2\over M_1^2\right)+
    g\left(M_i^2\over M_1^2\right)\right]\label{cpa}\; ;
  \eeqa
  here $f$ is the contribution from the one-loop vertex correction, 
  \beq
    f(x)=\sqrt{x}\left[1-(1+x)\ln\left(1+x\over x\right)\right]\;,
  \eeq
  and $g$ denotes the contribution from the one-loop self energy
  \cite{fps95,crv96,bp98}, which can be reliably calculated in perturbation
  theory for sufficiently large mass splittings, i.e.,
  $|M_i-M_1|\gg|\G_i-\G_1|$,
  \beq
    g(x)={\sqrt{x}\over1-x}\;.
  \eeq
For $M_1 \ll M_2, M_3$ one obtains
\beq\label{eps}
    \ve_1 \simeq -{3\over8\pi}\;{1\over\left(h_\n h_\n^\dg\right)_{11}}
    \sum_{i=2,3}\mbox{Im}\left[\left(h_\n h_\n^\dg\right)_{1i}^2\right]
    {M_1\over M_i}\; .
  \eeq
In the case of mass differences of order the decay widths 
one expects an enhancement from the self-energy
  contribution \cite{pil99}, although the influence of the thermal bath on this
  effect is presently unclear.

  The CP asymmetry (\ref{cpa}) then leads to a lepton
  asymmetry \cite{kt90},
  \beq\label{basym}
    Y_L\ =\ {n_L-n_{\Bar{L}}\over s}\ =\ \k\ {\ve_1\over g_*}\;.
  \eeq
  Here the factor $\k<1$ represents the effect of washout
  processes. In order to determine $\k$ one has to solve the full
  Boltzmann equations \cite{lut92,plu97}. In the examples discussed
  in section \ref{examples} one obtains $\k\simeq 10^{-1}\ldots 10^{-3}$.

  In a complete analysis one also has to consider washout processes.
  Particularly important are $\D L=2$ lepton Higgs scatterings
  mediated by heavy neutrinos (cf.~fig.~\ref{fig_lept}) since
  cancellations between on-shell contributions to these scatterings
  and contributions from neutrino decays and inverse decays ensure
  that no asymmetry is generated in thermal equilibrium \cite{kw80,dz81}.

  Further, due to the large top-quark Yukawa coupling one has to take
  into account neutrino top-quark scatterings mediated by Higgs bosons
  \cite{lut92,plu97}. As we will see in section \ref{qualitative},
  these processes are of crucial importance for leptogenesis, since
  they can create a thermal population of heavy neutrinos at high
  temperatures $T>M_1$. As the CP asymmetry can
  be interpreted as a mean lepton asymmetry produced per neutrino
  decay, the requested baryon asymmetry can only be generated if
  the neutrinos are sufficiently numerous before decaying.

Various extensions of the standard model have been considered in early
studies of leptogenesis \cite{lpy86}-\cite{fp98}. In particular, it is
intriguing that in the simple case of hierarchical heavy neutrino masses
the observed value of the baryon asymmetry is obtained without any fine tuning
of parameters if $B-L$ is broken at the unification scale, 
$\L_{GUT} \sim 10^{16}$~GeV. The corresponding light neutrino masses
are very small, i.e., 
$m_{\n_2} \sim 3\cdot 10^{-3}$~eV, as preferred by the MSW explanation of
the solar neutrino deficit, and $m_{\n_3} \sim 0.1$~eV \cite{bp96}. 
Such small neutrino masses are also consistent with the atmospheric neutrino
anomaly \cite{atm98}, which implies a small mass $m_{\n_3}$ in the
case of hierarchical neutrino masses. This fact  gave rise to a renewed 
interest in leptogenesis in recent years, and a number of interesting models 
have been suggested \cite{ars98}-\cite{no00}.

\subsection[Neutrino masses and mixings]{Neutrino masses and mixings \protect\label{nu_beg}}

Leptogenesis relates the cosmological baryon asymmetry and neutrino masses
and mixings. The predicted value of the baryon asymmetry
depends on the CP asymmetry (\ref{cpa}) which is determined by the Dirac and 
the Majorana neutrino mass matrices. Depending on the neutrino mass hierarchy 
and the size of the mixing angles the CP asymmetry can vary over many orders of
magnitude. It is therefore important to see whether patterns of neutrino
masses \cite{lv98} motivated by other considerations are consistent with 
leptogenesis. In the following we shall consider two examples.

An attractive framework to explain the observed mass hierarchies of quarks
and charged leptons is the Froggatt-Nielsen mechanism \cite{fn79} based
on a spontaneously broken U(1)$_F$ generation symmetry. The Yukawa couplings 
arise from non-renormalizable interactions after a gauge singlet field $\F$ 
acquires a vacuum expectation value,
\beq
h_{ij} = g_{ij} \left({\VEV\F\over \L}\right)^{Q_i + Q_j}\;.
\eeq
Here $g_{ij}$ are couplings $\co(1)$ and $Q_i$ are the U(1) charges of the
various fermions, with $Q_{\F}=-1$. The interaction scale $\L$ is
usually chosen to be very large, $\L > \L_{GUT}$. In the following we shall
discuss two different realizations of this idea which are motivated by
the atmospheric neutrino anomaly \cite{atm98}. Both
scenarios have a large $\n_\m -\n_\t$ mixing angle. They differ, however,
by the symmetry structure and by the size of the parameter $\e$ which
characterizes the flavour mixing.

\subsubsection[$SU(5)\times U(1)_F$]{$SU(5)\times U(1)_F$ \protect\label{su5masses}}
 
This symmetry has been considered by a number of authors.
Particularly interesting is the case with a non-parallel family structure
where the chiral $U(1)_F$ charges are different for the $\bf 5^*$-plets
and the $\bf 10$-plets of the same family \cite{ys99}-\cite{by99}. An
example of possible charges $Q_i$ is given in table~1.

\begin{table}[b]
\begin{center}
\begin{tabular}{c|ccccccccc}\hline \hline
$\j_i$       & $ e^c_{R3}$ & $ e^c_{R2}$  & $ e^c_{R1}$  & $ l_{L3}$    & 
$ l_{L2}$    & $ l_{L1}$   & $ \n^c_{R3}$ & $ \n^c_{R2}$ & $ \n^c_{R1}$ 
\\\hline
$Q_i$  & 0 & 1 & 2 & $0$ & $0$ & $1$ & 0 & $1$ & $2$ \\ \hline\hline
\end{tabular}
\end{center}
\caption{{\it Chiral charges of charged and neutral leptons with
   $SU(5)\times U(1)_F$ symmetry} \protect\cite{by99}.}
\end{table}

The assignment of the same charge to the lepton doublets of the second and 
third generation leads to a neutrino mass matrix of the form
\cite{ys99,ram99}, 
\beq\label{matrix}
m_{\n_{ij}} \sim \left(\begin{array}{ccc}
    \e^2  & \e  & \e \\
    \e  & \; 1 \; & 1 \\
    \e  &  1  & 1 
    \end{array}\right) {v_2^2\over \VEV R}\;.
\eeq
This structure immediately yields a large $\n_\m -\n_\t$ mixing angle. The
phenomenology of neutrino oscillations depends on the unspecified coefficients
$\co(1)$. The parameter $\e$ which gives the flavour mixing is chosen to be
\beq\label{exp1}
{\VEV\F\over\L} = \e  \sim {1\over 17}\;. 
\eeq
The three Yukawa matrices for the leptons have the structure,
\beq\label{yuk1}
h_e, \ 
h_{\n}  \sim\ \left(\begin{array}{ccc}
    \e^3 & \e^2 & \e^2 \\
    \e^2 &\;  \e \;   & \e   \\
    \e   & 1    & 1
    \end{array}\right) \;, \quad
h_{r}  \sim\ \left(\begin{array}{ccc}
    \e^4 & \e^3 & \e^2 \\
    \e^3 &\;  \e^2 \; & \e   \\
    \e^2 & \e   & 1
    \end{array}\right) \;.
\eeq
Note, that $h_e$ and $h_\n$ have the same, non-symmetric structure.
One easily verifies that the mass ratios for charged leptons, heavy and
light Majorana neutrinos are given by 
\beqa
\qquad\quad 
m_e : m_\m : m_\t \sim \e^3 : \e : 1\;, &\quad&
M_1 : M_2  : M_3  \sim \e^4 : \e^2 : 1\;,\\
m_1 : m_2  : m_3  &\sim& \e^2 : 1 : 1\;.
\eeqa
The masses of the two eigenstates $\n_2$ and $\n_3$ depend on unspecified 
factors of order one, and may differ by an order of magnitude 
\cite{ilr98,vis98}. They can therefore be consistent with the mass differences 
$\D m^2_{\n_1 \n_2}\simeq 4\cdot 10^{-6} - 1\cdot 10^{-5}$~eV$^2$ \cite{hl97}
inferred from the MSW solution of the solar neutrino problem \cite{msw86} and 
$\D m^2_{\n_2 \n_3}\simeq (5\cdot 10^{-4}-6\cdot 10^{-3})$~eV$^2$ associated 
with the atmospheric neutrino deficit \cite{atm98}. For numerical estimates 
we shall use the average of the neutrino masses of the second and third family,
$\Bar{m}_\n=(m_{\n_2}m_{\n_3})^{1/2} \sim 10^{-2}$~eV. 

\begin{table}
\begin{center}
\begin{tabular}{c|ccccccccc}
\hline \hline
$\j_i$       & $ e^c_{R3}$ & $ e^c_{R2}$  & $ e^c_{R1}$  & $ l_{L3}$    & 
$ l_{L2}$    & $ l_{L1}$   & $ \n^c_{R3}$ & $ \n^c_{R2}$ & $ \n^c_{R1}$ 
\\\hline
$Q_i$  & 0 & ${1\over 2}$ & ${5\over 2}$ & $0$ & ${1\over 2}$ & ${5\over 2}$ 
& 0 & ${1\over 2}$ & ${5\over 2}$ \\ \hline\hline
\end{tabular}
\end{center}
\caption{{\it Chiral charges of charged and neutral leptons with
$SU(3)_c \times SU(3)_L \times SU(3)_R \times U(1)_F$ symmetry}
\protect\cite{lr99}.}
\end{table}

The choice of the charges in table~1 corresponds to large Yukawa couplings
of the third generation. For the mass of the heaviest Majorana neutrino
one finds
\beq
M_3\ \sim\ {v_2^2\over\Bar{m}_\n}\ \sim\ 10^{15}\ \mbox{GeV}\;.
\eeq
Since $h_{r33}$ and the gauge coupling of $U(1)_{B-L}$ are ${\cal O}(1)$,
this implies that $B-L$ is broken at the unification scale $\L_{GUT}$.

\subsubsection[$SU(3)_c \times SU(3)_L \times SU(3)_R \times U(1)_F$]{$SU(3)_c \times SU(3)_L \times SU(3)_R \times U(1)_F$ \protect\label{nu_end}}

This symmetry arises in unified theories based on the gauge group $E_6$.
The leptons $e_R^c$, $l_L$ and $\n_R^c$ are contained in a single
$(1,3,\bar{3})$ representation. Hence, all leptons of the same generation
have the same $U(1)_F$ charge and all leptonic Yukawa matrices are
symmetric. Masses and mixings of quarks and charged leptons can be
successfully described by using the charges given in table~2 \cite{lr99}. 
Clearly, the three Yukawa matrices have the same structure\footnote{Note,
that with respect to ref.~\cite{lr99}, $\e$ and $\Bar{\e}$ have been
interchanged.},
\beq\label{yuk2}
h_e,\ h_r  \sim\ \left(\begin{array}{ccc}
    \e^5        & \e^3        & \e^{5/2} \\
    \e^3        & \e          & \e^{1/2} \\
    \e^{5/2} \; & \e^{1/2} \; & 1
    \end{array}\right) \;, \quad
h_{\n}  \sim\ \left(\begin{array}{ccc}
    \bar\e^5        &\bar\e^3        & \bar\e^{5/2} \\
    \bar\e^3        &\bar\e          & \bar\e^{1/2} \\
    \bar\e^{5/2} \; &\bar\e^{1/2} \; & 1
    \end{array}\right) \;,
\eeq
but the expansion parameter in $h_{\n}$ may be different from the one in
$h_e$ and $h_r$. From the quark masses, which also contain $\e$ and
$\bar{\e}$, one infers $\bar{\e} \simeq \e^2$ \cite{lr99}.

From eq.~(\ref{yuk2}) one obtains for the masses of charged leptons,
light and heavy Majorana neutrinos,
\beq
m_e : m_\m : m_\t\ \sim\ M_1 : M_2  : M_3\ \sim \e^5 : \e : 1\;, 
\eeq
\beq
m_1 : m_2 : m_3 \ \sim\ \e^{15} : \e^3 : 1\;.
\eeq
Like in the example with $SU(5)\times U(1)_F$ symmetry, the mass of the
heaviest Majorana neutrino,
\beq
M_3 \sim {v_2^2\over m_3} \sim 10^{15}\;\mbox{GeV} \;,
\eeq
implies that $B-L$ is broken at the unification scale $\L_{GUT}$.

The $\n_\m-\n_\t$ mixing angle is related to the mixing of
the charged leptons of the second and third generation \cite{lr99},
\beq
\sin{\Q_{\m\t}} \sim \sqrt{\e} + \e \;.
\eeq
This requires large flavour mixing,
\beq\label{exp2}
\left({\VEV\F\over\L}\right)^{1/2} = \sqrt{\e}  \sim {1\over 2}\;. 
\eeq 
In view of the unknown coefficients $\co(1)$ the corresponding mixing angle 
$\sin{\Q_{\m\t}} \sim 0.7$ is consistent with the interpretation of the 
atmospheric neutrino anomaly as $\n_\m-\n_\t$ oscillation.

It is very instructive to compare the two scenarios of lepton masses and
mixings described above. In the first case, the large $\n_\m-\n_\t$
mixing angle follows from a non-parallel flavour symmetry. The parameter $\e$,
which characterizes the flavour mixing, is small. In the second case, the
large $\n_\m-\n_\t$ mixing angle is a consequence of the large flavour
mixing $\e$. The $U(1)_F$ charges of all leptons are the same, i.e., one
has a parallel family structure. Also the mass hierarchies, given in terms
of $\e$, are rather different. This illustrates that the separation into
a flavour mixing parameter $\e$ and coefficients $\co(1)$ is far from
unique. It is therefore important to study other observables which
depend on the lepton mass matrices. This includes lepton-flavour changing
processes and, in particular, also the cosmological baryon asymmetry. 

\newpage

\section{Boltzmann equations and scattering processes}

\subsection{Boltzmann equations}

A full quantum mechanical description of baryogenesis has to take into
account the interplay of all processes in the plasma, i.e. decays, 
inverse decays and scattering processes together with the time evolution
of the system. Such a quantum mechanical treatment may be based either
on the time evolution of the density matrix \cite{jmy98} or on the 
Kadanoff-Baym equations for the Green functions of the system. In the
latter case a systematic perturbative expansion has recently been obtained
which starts from a set of Boltzmann equations for distribution functions 
\cite{bf00}. So far, however, all detailed studies of baryogenesis are
based on the Boltzmann equations for number densities. In this section we
therefore briefly review the basic ingredients of this approach.
In particular it is assumed that between scatterings the particles move freely
in the gravitational field of the expanding universe and that the interactions
are described by quantum field theory at zero temperature.

  Consider first a point-particle with mass $m\ge0$ moving freely in a
  gravitational field. Its coordinates in phase space $x^{\mu}$,
  $p^{\mu}$ obey the geodesic equations of motion \cite{ehlers}
  \beqa
     &&{\mbox{d}p^{\m}\over\mbox{d}\t}+\G^{\m}_{\mbox{ }\n\a}
     p^{\n}p^{\a}=0\label{A2}\;,\\[1ex]
     &&{\mbox{d}x^{\m}\over\mbox{d}\t}=p^{\m}\;.
  \eeqa
  If $p^{\mu}$ is the 4-momentum of the particle, the affine parameter
  $\tau$ is uniquely determined, except for its origin. For a massive
  particle, $m>0$, $s=m\tau$ is the proper time.
  With the Liouville operator
  \beq
    L=p^{\a}{\partial\over\partial x^{\a}}-\G^{\a}_
    {\mbox{ }\b\g}p^{\b}p^{\g}{\partial\over\partial p^{\a}}
    \label{liouville}
  \eeq
  the geodesic equations can be written as
  \beqa
    &&{\mbox{d}p^{\m}\over\mbox{d}\t}=L\left[p^{\m}\right]\;,
    \label{15a}\\[1ex]
    &&{\mbox{d}x^{\m}\over\mbox{d}\t}=L\left[x^{\m}\right]\;.
    \label{15b}
  \eeqa
  Consider now a gas of non-interacting particles, i.e.\ of particles
  whose motion is determined by the equations (\ref{15a}) and
  (\ref{15b}). Due to the absence of interactions, the phase space
  density $f_{\j}(x,p)$ has to be constant, i.e.\ its total derivative
  with respect to the parameter $\tau$ has to vanish,
  \beq
    {\mbox{d}f_{\j}(x,p)\over\mbox{d}\t}=0\;.
    \label{boltz1}
  \eeq
  If we restrict the Liouville operator (\ref{liouville}) to the
  mass-shell of these particles,
  \beq
    L_m=p^{\a}{\partial\over\partial x^{\a}}-\G^i_{\mbox{ }\b\g}
    p^{\b}p^{\g}{\partial\over\partial p^i}\;,
    \label{liouville_m}
  \eeq
  where $i$ runs over spatial indices 1 to 3, the collisionless
  Boltzmann equation (\ref{boltz1}) reads \cite{ehlers}
  \beq
    L_m\left[f_{\j}(x,p)\right]=0\;. \label{17}
  \eeq
  In a spatially homogeneous and isotropic universe, described by the
  Robertson-Walker metric, the phase space density $f_{\j}$ can only
  be a function of $t$ and $\abs{\vec{p}_{\j}}$. The Boltzmann
  equation is then given by
  \beq\label{freeboltz}
    L_m\left[f_{\j}\right]=E_{\j}{\partial f_{\j}\over\partial t}-
    H\left|\vec{p}_{\j}\right|^2{\partial f_{\j}\over\partial E_{\j}}=0\;,
  \eeq
  where $E_{\j}=\sqrt{\vec{p}_{\j}^2+m_{\j}^2}$ and $H$ is the Hubble
  parameter. The equilibrium phase space density $f_{eq}=\exp{(-E_{\j}/T)}$ 
  solves this
  equation only in the limit $m_{\j}\to0$ or $m_{\j}\to\infty$, i.e.,
  only extremely relativistic or non-relativistic particles can be in
  thermal equilibrium in a Roberston-Walker universe. This is due to
  the fact that the Robertson-Walker metric has no timelike spatially 
  constant Killing vector \cite{bern}. Other solutions of (\ref{freeboltz})
  are the Bose-Einstein and the Fermi-Dirac distributions of massless
  particles. 
  
  However, one can come close to thermal equilibrium by including
  interactions, which are described by a collision term
  $C\left[f_{\j}\right]$ in the Boltzmann equation:
  \beq
    E_{\j}{\partial f_{\j}\over\partial t}-H\left|\vec{p}_{\j}\right|^2
    {\partial f_{\j}\over\partial E_{\j}}=C\left[f_{\j}\right]\;.
  \eeq
  Integrating over the phase space element 
  \beq
    \dd\tilde{p}_{\j}={\dd^3p_{\j}\over(2\p)^32E_{\j}} \label{1_25}
  \eeq
  yields
  \beq
    \dot{n}_{\j}+3Hn_{\j}={g_{\j}\over(2\p)^3}\int{\mbox{d}^3p_{\j}\over
    E_{\j}}\,C\left[f_{\j}\right]\;,
  \eeq
  where we have made use of the mass shell condition and the spatial
  isotropy, and $n_{\j}$ is the number density.

  The collision term counts the number of collisions a particle $\j$
  undergoes in a time and volume element. For the process
  $\j+a+b+\ldots\to i+j+\ldots$ it is given by \cite{wag,kw80,lut92}:
     \beqa
     &&\hspace{-1cm}\g\left(\j+a+b+\ldots\to i+j+\ldots\right):=
     -{g_{\j}\over(2\p)^3}\int{\mbox{d}^3p_{\j}\over E_{\j}}\,
     C\left[f_{\j}\right]\NO\\[1ex]
     &&=\int\dd\tilde{p}_{\j}\dd\tilde{p}_a\dd\tilde{p}_b
     \ldots\dd\tilde{p}_i\dd\tilde{p}_j\ldots(2\p)^4\d^4
     \left(p_{\j}+p_a+p_b+\ldots-p_i-p_j-\ldots\right)\NO\\[1ex]
     &&\quad\times\left|\ampl{\j+a+b+\ldots}{i+j+\ldots}\right|^2
     f_{\j}f_af_b\ldots\left(1\pm f_i\right)\left(1\pm f_j\right)\ldots\;,
     \label{1_24}
     \eeqa
  where the squared matrix element $\abs{\ampl{\j+a+b+\ldots}
  {i+j+\ldots}}^2$ has to be summed over internal degrees of freedom
  of incoming and outgoing particles. A symmetry factor $1/n!$ has to
  be included if there are $n$ identical incoming or outgoing
  particles.

  In a dilute gas the factors $\left(1\pm f_i\right)$, where the upper
  (lower) sign refers to bosons (fermions), can be neglected, and we
  only have to consider decays $\j\to i+j+\ldots$, two particle
  scatterings $\j+a\to i+j+\ldots$, and the corresponding back
  reactions. Then the Boltzmann equation for $n_{\j}$ reads:
  \beqa
    \hspace{-2cm}\dot{n}_{\j}+3Hn_{\j}&=&-\sum\limits_{i,j,\ldots}
    \left[\g\left(\j\to i+j+\ldots\right)-
    \g\left(i+j+\ldots\to\j\right)\right]\NO\\[1ex]
    &&-\sum\limits_{a,i,j,\ldots}\left[
    \g\left(\j+a\to i+j+\ldots\right)-
    \g\left(i+j+\ldots\to\j+a\right)\right]\;.
    \label{1_26}
  \eeqa
  Let us consider the decay term first. From the definition
  (\ref{1_24}) it follows that
  \beqa
     &&\hspace{-1.5cm}\sum\limits_{i,j,\ldots}\left[\g\left(\j\to 
     i+j+\ldots\right)-\g\left(i+j+\ldots\to\j\right)\right]\NO\\[1ex]
     &&=\sum\limits_{i,j,\ldots}\int\dd\tilde{p}_{\j}\,\dd\tilde{p}_i\,
     \dd\tilde{p}_j\ldots(2\p)^4\d^4\left(p_{\j}-p_i-p_j-\ldots\right)
     \times\NO\\[1ex]
     &&\quad\left[\abs{\ampl{\j}{i+j+\ldots}}^2f_{\j}
     -\abs{\ampl{i+j+\ldots}{\j}}^2f_if_j\ldots\right]\;.
  \eeqa
  In thermal equilibrium we have $f_if_j\ldots=f_{\j}$ because of
  energy conservation. Hence, due to the unitarity of the $S$-matrix,
  this decay term vanishes in thermal equilibrium. The same holds
  true for scattering processes, even if quantum corrections are
  included \cite{kw80,wag,bf00}. 
   
  Further, one can distinguish elastic and inelastic scatterings.
  Elastic scatterings only affect the phase space densities of the
  particles but not the number densities, whereas inelastic
  scatterings do change the number densities. If elastic scatterings
  do occur at a higher rate than inelastic scatterings one can assume
  kinetic equilibrium, i.e., the phase space density is
  \beq
    f_{\j}(E_{\j},T)={n_{\j}\over n_{\j}^{eq}}\mbox{e}^{-E_{\j}/T}\;.
  \eeq
  The Boltzmann equation (\ref{1_26}) then takes the form
  \beqa
     \dot{n}_{\j}+3Hn_{\j}&=& \label{1_28}                \\
&&\hspace{-3cm}-\sum\limits_{i,j,\ldots}
     \left[{n_{\j}\over n_{\j}^{eq}}\,\g^{eq}\left(\j\to i+j+\ldots\right)-
     {n_in_j\ldots\over n_i^{eq}n_j^{eq}\ldots}
     \,\g^{eq}\left(i+j+\ldots\to\j\right)\right]\NO\\[1ex]
     &&\hspace{-3cm}-\sum\limits_{a,i,j,\ldots}\left[{n_{\j}n_a\over
     n_{\j}^{eq}n_a^{eq}}\,\g^{eq}\left(\j+a\to i+j+\ldots\right)-
     {n_in_j\ldots\over n_i^{eq}n_j^{eq}\ldots}
     \,\g^{eq}\left(i+j+\ldots\to\j+a\right)\right]\; ,\NO
  \eeqa
  where $\g^{eq}$ is the space time density of a given process in
  thermal equilibrium. Note, that elastic scatterings no longer
  contribute to the evolution of $n_{\j}$.

  It is convenient to replace the particle density $n_{\j}$ by a
  quantity which is not affected by the expansion of the
  universe. Consider the number of particles in a comoving volume
  element, i.e., the ratio of particle density and entropy density $s$,
  \beq
     Y_{\j} \equiv {n_{\j}\over s}\;.
  \eeq
  If the universe expands isentropically,
  \beq
     \dot{n}_{\j}+3Hn_{\j}=s\dot{Y}_{\j}\;.\label{1_31}
  \eeq
  Hence, $Y_{\j}$ is not affected by the expansion, it can only be
  changed by interactions. Further, it is useful to transform to the
  dimensionless evolution variable $z=m_{\j}/ T$. In a radiation
  dominated universe the Boltzmann equation (\ref{1_28}) then finally
  reads
  \beqa
      {sH\left(m_{\j}\right)\over z} {\mbox{d}Y_{\j}\over\mbox{d}z}&=&  \\
&&\hspace{-2.5cm}\sum\limits_{i,j,\ldots}
     \left[{Y_{\j}\over Y_{\j}^{eq}}\,\g^{eq}\left(\j\to i+j+\ldots\right)-
     {Y_iY_j\ldots\over Y_i^{eq}Y_j^{eq}\ldots}
     \,\g^{eq}\left(i+j+\ldots\to\j\right)\right]\NO\\[1ex]
     &&\hspace{-3cm}-\sum\limits_{a,i,j,\ldots}\left[{Y_{\j}Y_a\over
     Y_{\j}^{eq}Y_a^{eq}}\,\g^{eq}\left(\j+a\to i+j+\ldots\right)-
     {Y_iY_j\ldots\over Y_i^{eq}Y_j^{eq}\ldots}
     \,\g^{eq}\left(i+j+\ldots\to\j+a\right)\right]\;,\NO
     \label{1_34}
  \eeqa
  where $H(m_{\j})$ is the Hubble parameter at $T=m_{\j}$.

\subsection{Reaction densities}

  The reaction densities in thermal equilibrium $\g^{eq}$ can be
  further simplified. The decay width of a particle with energy $E$ is
  \beq
     \tilde{\G}={m\over E}\tilde{\G}_{rs}\;,\label{tildeg}
  \eeq
  where $\tilde{\G}_{rs}$ denotes the decay width in the rest
  system. The reaction density for the decay is then given by
  \beq
    \g_D^{eq}=
    n^{eq}_{\j}{\mbox{K}_1(z)\over\mbox{K}_2(z)}
    \,\tilde{\G}_{rs}\;;
    \label{1_39}
  \eeq
  here the ratio of Bessel functions is the thermal average of the
  time dilatation factor $m/E$ in eq.~(\ref{tildeg}).

  The decay rate $\G_D$ for the particle $\j$ is defined as the number
  of decays per time element, i.e., it is given by the ratio of
  $\g_D^{eq}$ and the number density $n_{\j}^{eq}$:
  \beq
    \G_D={\mbox{K}_1(z)\over\mbox{K}_2(z)}
    \,\tilde{\G}_{rs}\;.
  \eeq
  In the low and high temperature limits the decay rate is given
  by \cite{kw80}: 
  \beq
     \G_D\approx\left\{\begin{array}{l@{\quad}l}
     \displaystyle \tilde{\G}_{rs}&,\quad T\ll m_{\j}\\[2ex]
     \displaystyle {m_{\j}\over2T}\tilde{\G}_{rs}&,\quad T\gg m_{\j}
     \end{array}\right.
     \label{1_41}
  \eeq
  i.e., due to time dilatation, decays are suppressed at high
  temperatures.

  In the Boltzmann equation (\ref{1_28}) the expansion term $3Hn_{\j}$
  is responsible for deviations from thermal equilibrium, whereas
  interaction terms try to bring the system into equilibrium. Hence,
  the ratio of reaction rates $\G$ to the Hubble parameter $H$ is a
  measure for the effectiveness of the interactions. If the reaction
  rates are too small the system will be driven out of equilibrium by
  the expansion of the universe.

  For inverse decays we know from energy conservation that
  $f_i^{eq}f_j^{eq}\ldots=f_{\j}^{eq}$. If we further neglect possible
  CP violating effects we have
  \beq
    \ampl{\j}{i+j+\ldots}=\ampl{i+j+\ldots}{\j}\;,
  \eeq
  i.e., in thermal equilibrium, the reaction densities for decays and
  inverse decays are equal. If the decay products are massless, the
  reaction rate for inverse decays is \cite{kt90}:
  \beq
    \G_{ID}={n_{\j}^{eq}\over n_{\g}^{eq}}\,\G_{D}\;,
  \eeq
  where $n_{\g}^{eq}=(g/\p^2)T^3$ is the number density of a massless
  particle species with $g$ internal degrees of freedom in thermal
  equilibrium. In the limit of low and high temperatures we now have
  \beq
    \G_{ID}\approx\left\{\begin{array}{l@{\quad}l}
    \displaystyle \sqrt{\p}\left({m_{\j}\over2T}\right)^{3/2}
    \mbox{e}^{-m_{\j}/T}\,\tilde{\G}_{rs}&,\quad T\ll m_{\j}\\[2ex]
    \displaystyle {m_{\j}\over2T}\tilde{\G}_{rs}&,\quad T\gg m_{\j}\;.
    \end{array}\right.\label{1_44}
  \eeq

  For a two particle scattering process one obtains \cite{lut92}
  \beq
     \g^{eq}({\j}+a\to i+j+\ldots)
     ={T\over64\p^4}\int\limits_{\left(m_{\j}+m_a\right)^2}^{\infty}
     \hspace{-0.5cm}\dd s\,\hat{\s}(s)\,\sqrt{s}\,
     \mbox{K}_1\left({\sqrt{s}\over T}\right)\;,\label{1_50}
  \eeq
  where $\hat{\s}(s)$ is the reduced cross section
  \beq
    \hat{\s}(s) 
     ={2\l\left(s,m_{\j}^2,m_a^2\right)\over s}\,\s(s)\label{1_49}\;.
  \eeq
  Here, $\s(s)$ is the usual cross section as a function of the
  squared centre of mass energy $s$ and $\l$ is a flux factor
  \beq
    \l(x^2,y^2,z^2)=\sqrt{\left[x^2-(y+z)^2\right]
            \left[x^2-(y-z)^2\right]}\;.
  \eeq
  If $\abs{\vec{v}}$ denotes the relative velocity of the two particles,
  \beq
     \abs{\vec{v}}={\sqrt{\left(p_{\j}\cdot p_a\right)^2-
     m_{\j}^2m_a^2}\over E_{\j}\,E_a}\;,
  \eeq
  the reaction density for the scattering can also be written as
  \beq
    \g^{eq}(\j+a\to i+j+\ldots)=n_{\j}^{eq}\,n_a^{eq}\,
    \VEV{\s\,\abs{\vec{v}}}\;.
  \eeq
In the following we shall consider processes with up to three particles in
the final state.

\section[Baryon asymmetry]{Baryon asymmetry \protect\label{examples}}

  We are now in a position to evaluate the baryon asymmetry for the patterns of
  neutrino masses discussed in section
  \ref{nu_beg}. Before doing that, however, it is useful
  to consider some general aspects of leptogenesis, which are
  independent of any particular parametrization chosen for the
  neutrino mass matrices. This supplements the qualitative
  discussion of section \ref{nudecay}.

\subsection[Delayed decay]{Delayed decay \protect\label{qualitative}}

\begin{figure}[t]
     \centerline{\epsfig{file=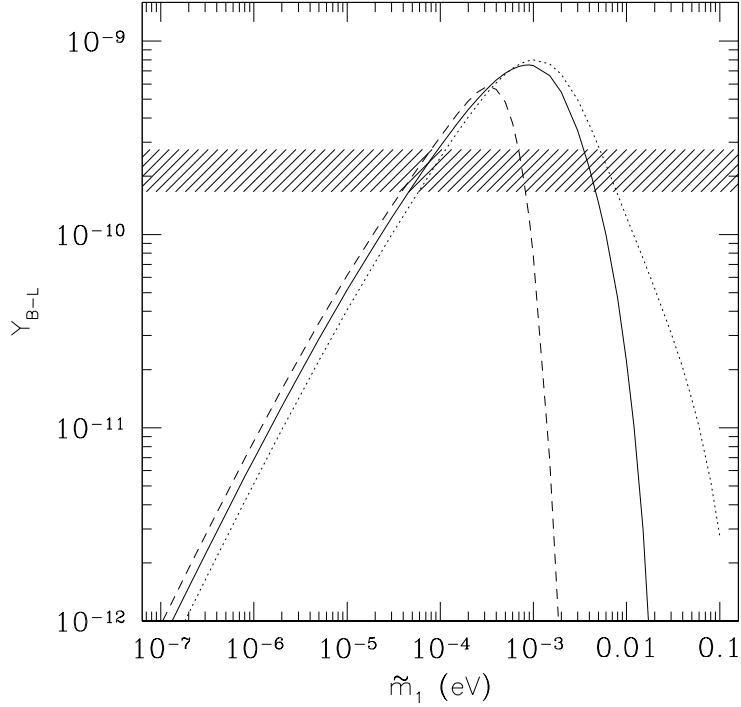,width=10cm}}
     \caption{\it Generated $B-L$ asymmetry as a function of
     $\wt{m}_1$ for $M_1=10^8\,$GeV (dotted line), $M_1=10^{10}\,$GeV 
     (solid line) and $M_1=10^{12}\,$GeV (dashed line). The shaded
     area shows the measured value for the asymmetry.
     \label{m1tilde_plot}}
  \end{figure}

  A characteristic feature of this baryogenesis scenario is that the
  generated asymmetry depends mostly on the mass parameter
  \beq
    \wt{m}_1\ =\ (h_\n h_\n^\dg)_{11} {v_2^2\over M_1}\;.
  \eeq
This can be seen in fig.~\ref{m1tilde_plot} where we have plotted the 
generated lepton
asymmetry as a function of $\wt{m}_1$ for three different heavy
neutrino masses $M_1=10^8\,$GeV, $10^{10}\,$GeV and $10^{12}\,$GeV.
Further, we have assumed a fixed CP asymmetry $\ve_1=-10^{-6}$ and a
fixed mass hierarchy for right-handed neutrinos,
$M_2^2=10^3\,M_1^2$, $M_3^2=10^6\,M_1^2$.

  First one sees that leptogenesis is only possible in a rather narrow
  range of $\wt{m}_1$, and that the washout processes, parametrized by
  the factor $\kappa$ in eq.~(\ref{basym}), are important over the whole 
parameter range, since $\kappa < 0.1$. If $\wt{m}_1$ is too low, the Yukawa
  interactions are too weak to produce a sufficient number of
  neutrinos at high temperatures, whereas for large $\wt{m}_1$ the
  washout processes are too strong and destroy any generated
  asymmetry.

  The asymmetry depends almost only on $\wt{m}_1$ for small
  $\wt{m}_1\,\ltap\,10^{-4}\,$eV, since then the generated asymmetry
  is determined by the number of neutrinos produced at high
  temperatures, i.e., on the strength of the processes in which a
  right-handed neutrino can be generated or annihilated. The dominant
  reactions are decays, inverse decays and scatterings with a top. In
  the Boltzmann equation (\ref{1_34}) the reaction densities for heavy
  neutrino production, $\g_{\scr PROD}$, all give contributions
  proportional to $\wt{m}_1$ at high temperatures $T>M_1$,
  \beq
    {-z\over sH(M_1)}\;\g_{\scr PROD}\propto\wt{m}_1\;.
    \label{propor}
  \eeq
  For large $\wt{m}_1\,\gtap\,10^{-4}\,$eV on the other hand, the
  neutrinos reach thermal equilibrium at high temperatures, i.e., the
  generated asymmetry depends mostly on the influence of the lepton
  number violating scatterings $\g_{\scr \D L = 2}$ at temperatures
  $T\ltap M_1$. In contrast to eq.~(\ref{propor}) the lepton number
  violating processes mediated by a heavy neutrino behave at low temperatures
   like
  \beq
    {-z\over sH(M_1)}\;\g_{\scr \D L = 2}\propto
    M_1\sum\limits_j\wt{m}_j^2\;,
    \label{propor2}
  \eeq
with
\beq  
\wt{m}_j=(h_\n h_\n^\dg)_{jj} {v_2^2\over M_j}\;,
  \eeq
where we have neglected interference terms.
One therefore expects that the generated
  asymmetry becomes smaller for growing neutrino mass $M_1$ and this
  is exactly what one observes in fig.~\ref{m1tilde_plot}.

  Eq.~(\ref{propor2}) can also explain the small dependence of the
  asymmetry on the heavy neutrino mass $M_1$ for
  $\wt{m}_1\,\ltap\,10^{-4}\;$eV. The inverse decay processes which
  take part in producing the neutrinos at high temperatures are CP
  violating, i.e., they generate a lepton asymmetry at high
  temperatures.  Due to the interplay of inverse decay processes and
  lepton number violating $2\to2$ scatterings this asymmetry has a
  different sign compared to the one generated in neutrino decays at
  low temperatures, i.e., the asymmetries will partially cancel each
  other.  This cancellation can only be avoided if the asymmetry
  generated at high temperatures is washed out before the neutrinos
  decay. At temperatures $T > M_j$ the lepton number violating scatterings
  behave like
  \beq
      {-z\over sH(M_1)}\;\g_{\scr \D L = 2}\propto
      M_1\sum\limits_j{M_j^2\over M_1^2}\;\wt{m}_j^2\;.
  \eeq
  Hence, for fixed heavy neutrino mass hierarchy, the wash-out
  processes are more efficient for larger neutrino masses, i.e., the
  final asymmetry should grow with the neutrino mass $M_1$. The
  finally generated asymmetry is not affected by the stronger wash-out
  processes, since for small $\wt{m}_1$ the neutrinos decay late,
  where one can neglect the lepton number violating scatterings.

\subsection{Leptogenesis}

We can now evaluate the baryon asymmetry for the two patterns of neutrino
mass matrices discussed in sections \ref{su5masses} and \ref{nu_end}.
Since for the Yukawa couplings only the powers in $\e$ are known, we will
also obtain the CP asymmetries and the corresponding baryon asymmetries
to leading order in $\e$, i.e., up to unknown factors ${\cal O}(1)$. 
Note, that for models with a $U(1)_F$ generation symmetry the baryon
asymmetry is `quantized', i.e., changing the $U(1)_F$ charges will change
the baryon asymmetry by powers of $\e$ \cite{by99}.

\subsubsection{$SU(5)\times U(1)_F$}

In this case one obtains from eqs.~(\ref{cpa}) and (\ref{yuk1}),
\beq
  \ve_1\ \sim\ {3\over 16\pi}\ \e^4\;.
  \label{su5epsilon}
\eeq 
From eq.~(\ref{basym}), $\e^2 \sim 1/300$ (\ref{exp1}) and $g_* \sim 100$ 
one then obtains the baryon asymmetry,
\beq\label{est1}
Y_B \sim \k\ 10^{-8}\;.
\eeq
For $\k \sim 0.1\ldots 0.01$ this is indeed the correct order of magnitude!
The baryogenesis temperature is given by the mass of the lightest of the
heavy Majorana neutrinos,
\beq
T_B \sim M_1 \sim \e^4 M_3 \sim 10^{10}\ \mbox{GeV}\;.
\eeq

\begin{figure}[t]
    \mbox{ }\hfill
    \epsfig{file=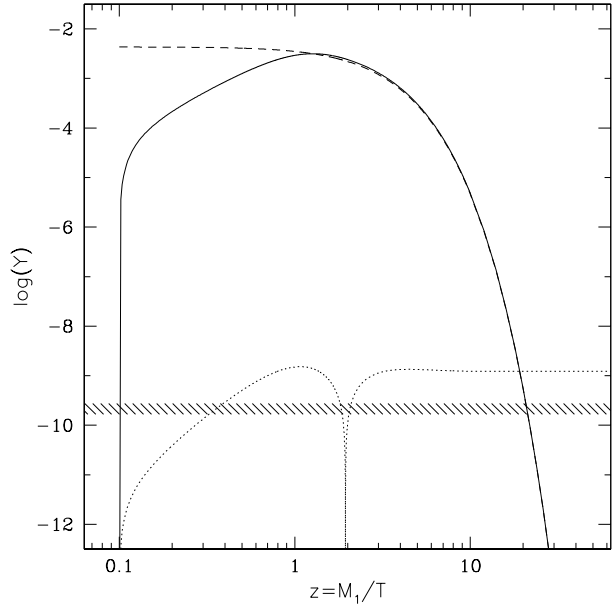,width=8.2cm}
    \hfill\mbox{ }
    \caption{\it Time evolution of the neutrino number density and the
     lepton asymmetry in the case of the $SU(5)\times U(1)_F$ symmetry. 
     The solid line shows the solution of the Boltzmann equation for the 
     right-handed neutrinos, while the corresponding equilibrium 
     distribution is represented by the dashed line.
     The absolute value of the lepton asymmetry $Y_L$ 
     is given by the dotted line and the hatched area shows the
     lepton asymmetry corresponding to the observed baryon asymmetry.
     \label{asyB}}
  \end{figure}

This set of parameters, where the CP asymmetry is given in terms of the mass 
hierarchy of the heavy neutrinos, has been studied in detail \cite{bp96}. 
The generated baryon asymmetry does not depend on the flavour mixing of the 
light neutrinos, in particular the $\n_\m-\n_\t$ mixing angle. 
The solution of the full Boltzmann equations is shown in fig.~\ref{asyB} 
for the non-supersymmetric case. The initial condition at a 
temperature $T \sim 10 M_1$ is chosen to be a state without heavy neutrinos. 
The Yukawa interactions are sufficient to bring the heavy neutrinos into 
thermal equilibrium. At temperatures $T\sim M_1$ this is followed by the usual 
out-of-equilibrium decays which lead to a non-vanishing baryon asymmetry. 
The final asymmetry agrees with the estimate (\ref{est1}) for $\k \sim 0.1$. 
The dip in fig.~7 is due to a change of sign in the lepton asymmetry at
$T \sim M_1$, as discussed in the previous section.

\subsubsection{$SU(3)_c\times SU(3)_L\times SU(3)_R\times U(1)_F$}

\begin{figure}[t]
    \mbox{ }\hfill
    \epsfig{file=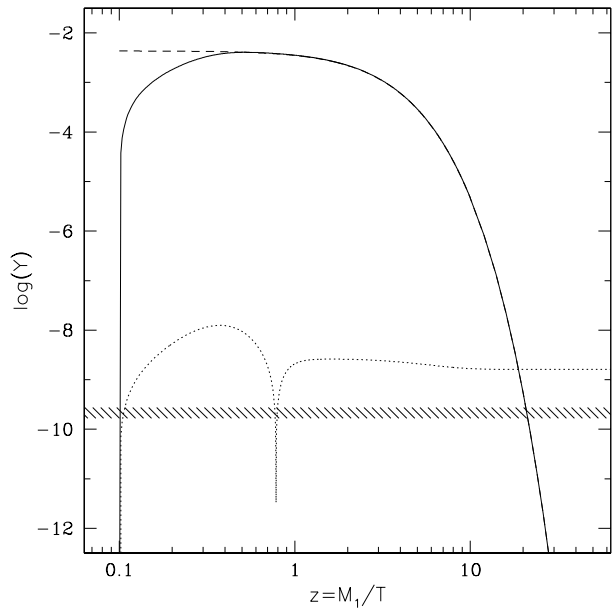,width=8.2cm}
    \hfill\mbox{ }
    \caption{\it Solution of the Boltzmann equations in the case of the
     $SU(3)_c\times SU(3)_L\times SU(3)_R\times$ $U(1)_F$ symmetry. 
     \label{asyLR}}
  \end{figure} 

In this case the neutrino Yukawa couplings (\ref{yuk2}) yield the CP asymmetry
\beq
\ve_1\ \sim\ {3\over 16\pi}\ \e^5\;,
\eeq 
which correspond to the baryon asymmetry (cf.~(\ref{basym}))
\beq\label{est2}
Y_B \sim \k\ 10^{-6}\;.
\eeq
Due to the large value of $\e$ the CP asymmetry is two orders of magnitude
larger than in the case with $SU(5)\times U(1)_F$ symmetry. However, 
washout processes are now also stronger. The solution of the Boltzmann 
equations is shown in fig.~\ref{asyLR}. The final asymmetry is again 
$Y_B \sim 10^{-9}$ 
which corresponds to $\k \sim 10^{-3}$. The baryogenesis temperature is
considerably larger than in the first case,
\beq
T_B \sim M_1 \sim \e^5 M_3 \sim 10^{12}\ \mbox{GeV}\;.
\eeq

The baryon asymmetry is largely determined by the parameter $\wt m_1$
defined in eq.~(\ref{ooeb}) \cite{plu97}. In the first example, one
has $\wt m_1 \sim \Bar m_\n$. In the second case one finds  
$\wt m_1 \sim m_3$. Since $\Bar m_\n$ and $m_3$ are rather similar
it is not too surprizing that the generated baryon asymmetry is about
the same in both cases.

\subsection{Supersymmetric leptogenesis}

  \begin{figure}
    \centerline{\input{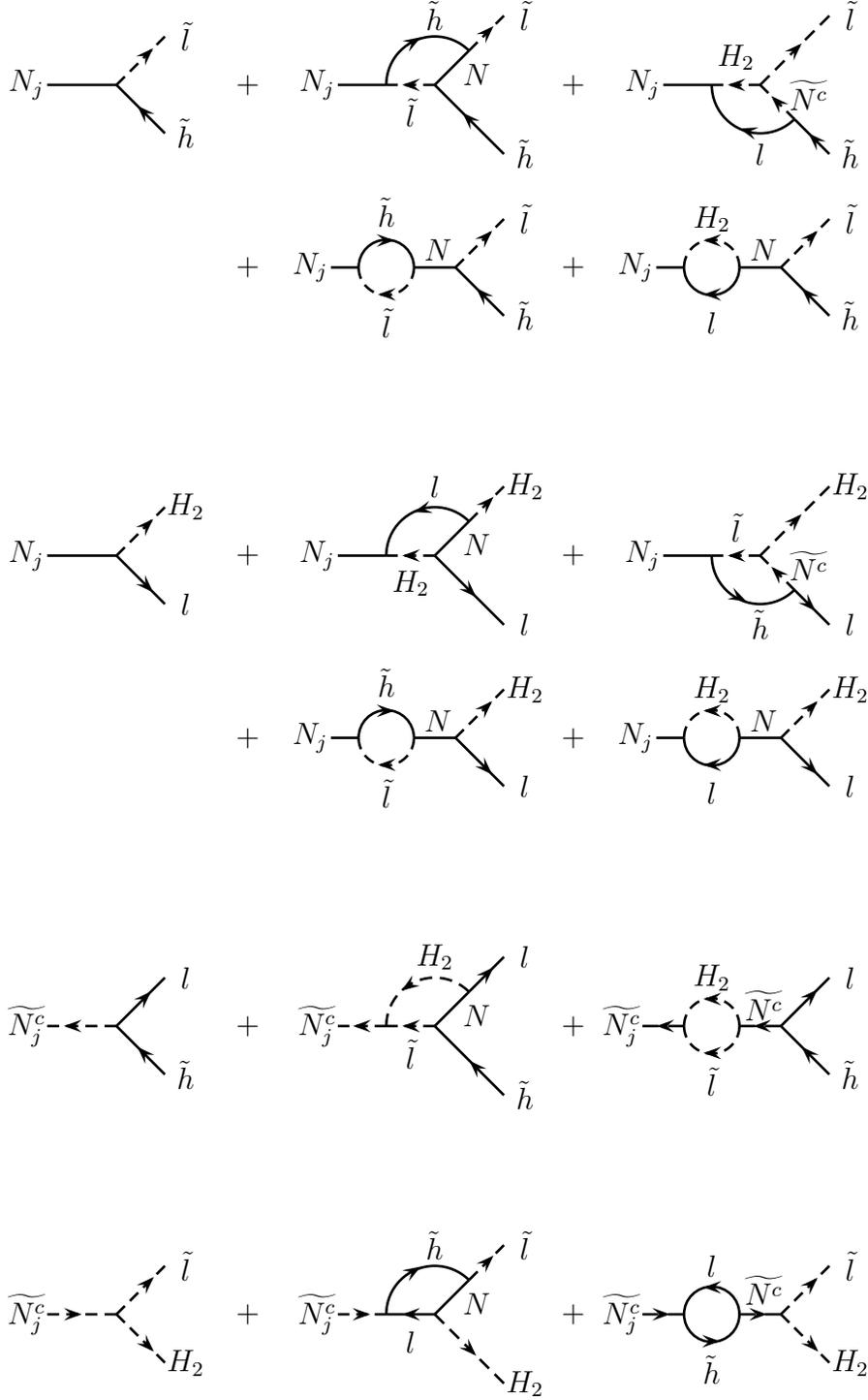}}
    \caption{\it Decay modes of the right-handed Majorana neutrinos
                 and their scalar partners in the supersymmetric
                 scenario. \label{susy_decay}}
  \end{figure}
  \begin{figure}[t]
     \centerline{\epsfig{file=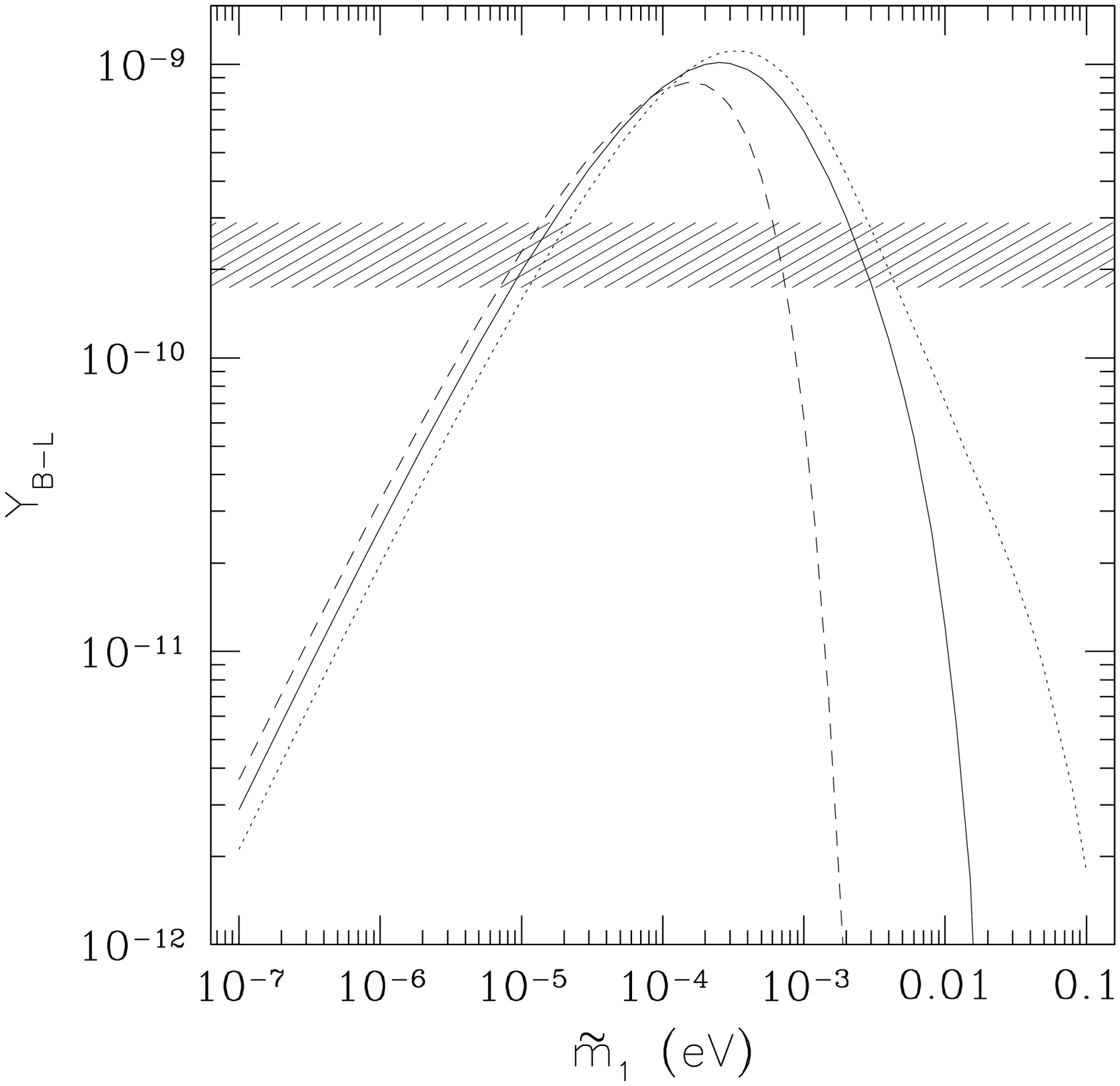,width=10cm}}
     \caption{\it Same as fig.~\ref{m1tilde_plot} for the 
      supersymmetric version of leptogenesis.
     \label{m1tilde_susy_plot}}
  \end{figure}
  Without an intermediate scale of symmetry breaking, the unification
  of gauge couplings appears to require low-energy supersymmetry.
  Therefore, we are going to briefly review supersymmetric 
  leptogenesis \cite{cdo93,crv96,plu98} in the following. As in the 
  non-supersymmetric case, a full analysis of the mechanism 
  including all the relevant scattering processes is necessary in 
  order to get a reliable relation between the input parameters and 
  the final asymmetry. 
  
  The supersymmetric generalization of the lagrangian (\ref{yuk}) is
  the superpotential
  \beq
    W = h_{eij}E_i^c L_j H_1 + h_{\n ij}N_i^c L_j H_2 +
        {1\over2}h_{rij} N_i^c N_j^c R \;,
  \eeq
  where, in the usual notation, $H_1$, $H_2$, $L$, $E^c$, $N^c$ and $R$ are
  chiral superfields describing spin-0 and spin-${1\over 2}$ fields.
  The basis for the lepton fields can be chosen as in the
  non-supersymmetric case.  

  The heavy neutrinos and their scalar partners can decay into
  various final states (cf.~fig.~\ref{susy_decay}). At tree level,
  the decay widths read,
  \beqa
    \G_{rs}\Big(N_1\to\wt{l}+\wt{h}^c\;\Big)
       =\G_{rs}\Big(N_1\to l+H_2\Big)
       &=&{1\over16\p}\;(h_\n h_\n^\dg)_{11} M_1\;,
         \label{decay1}\\[1ex]
    \G_{rs}\Big(\wt{N}_1^c\to\wt{l}+H_2\Big)
       =\G_{rs}\Big(\wt{N}_1\to l+\wt{h}^c\;\Big)
       &=&{1\over8\p}\;(h_\n h_\n^\dg)_{11} M_1\;.
        \label{decay2}
  \eeqa
  The CP asymmetry in each of the decay channels is given
  by \cite{crv96} 
  \beq
     \ve_1= -{1\over8\pi}\;{1\over\left(h_\n h_\n^\dg\right)_{11}}
    \sum_{i=2,3}\mbox{Im}\left[\left(h_\n h_\n^\dg\right)_{1i}^2\right]
    f\left(M_i^2\over M_1^2\right)\;,
  \eeq
  where
  \beq
    f(x)=\sqrt(x)\left[\ln\left({1+x\over x}\right)+{2\over x-1}\right]\;.
  \eeq
  It arises through interference of tree level and one-loop diagrams
  shown in fig.~\ref{susy_decay}.  In the case of a mass hierarchy,
  $M_j\gg M_i$, the CP asymmetry is twice as large as in the
  non-supersymmetric case.

\begin{figure}[t]
    \mbox{ }\hfill
    \epsfig{file=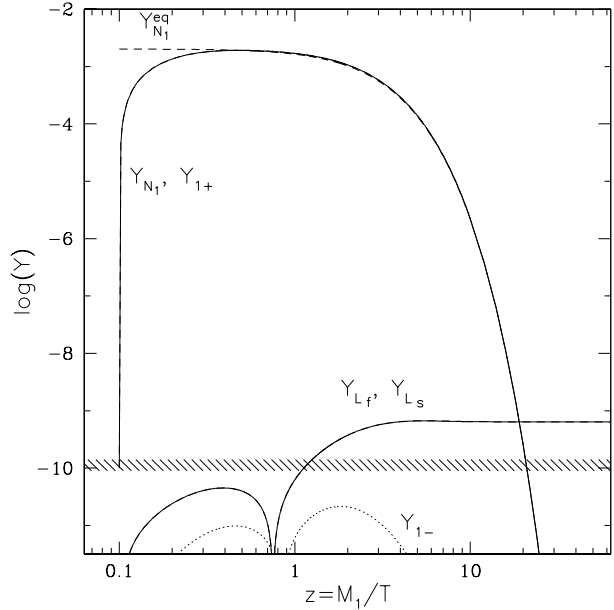,width=8.2cm}
    \hfill\mbox{ }
    \caption{\it Solution of the Boltzmann equations for supersymmetric
     leptogenesis for the parameters given by the $SU(5)\times U(1)_F$ 
     symmetry. $\Ynoneeq$ stands for the equilibrium distribution of
     neutrinos $N_1$, and $\Ynone$ and $\Ypone$ are the solutions for the
     neutrino number and the sum of scalar neutrino and scalar
     anti-neutrino number per comoving volume element,
     respectively. The generated lepton asymmetries in leptons and
     scalar leptons are denoted by $\YL$ and $\YLt$, whereas $\Ymone$
     stands for the difference of scalar neutrino and scalar
     anti-neutrino numbers. 
     \label{asyB_susy}}
  \end{figure}
  
  Like in the non-supersymmetric scenario lepton number violating
  scatterings mediated by heavy (s)neutrinos have to be included in a
  consistent analysis. In the supersymmetric case a large number of
processes contribute which can easily reduce the generated
  asymmetry by two orders of magnitude. Similarly, the large number of
(s)neutrino production processes makes leptogenesis possible for values
of $\wt m_1$ smaller than in the non-supersymmetric case \cite{plu98}. 

  In fig.~\ref{m1tilde_susy_plot} we have plotted the generated lepton
  asymmetry as function of $\wt{m}_1$ for three different values of
  $M_1$, where we have again assumed the hierarchy $M_2^2=10^3\;M_1^2$,
  $M_3^2=10^6\;M_1^2$ and the CP asymmetry $\ve_1=-10^{-6}$.
  Fig.~\ref{m1tilde_susy_plot} demonstrates that in the whole
  parameter range the generated asymmetry is significantly smaller than the
  value $4\cdot10^{-9}$ which one obtains, if one neglects lepton number 
violating scattering processes. Baryogenesis is possible in the range
  \beq
    10^{-5}\;\mbox{eV}\;\ltap\;\wt{m}_1\;\ltap\;
    5\cdot10^{-3}\;\mbox{eV}\;.
  \eeq

Comparing non-supersymmetric and supersymmetric leptogenesis one sees 
that the larger CP asymmetry and the additional contributions from the 
sneutrino decays in the supersymmetric scenario are compensated by the 
wash-out processes which are stronger than in the non-supersymmetric 
case. The final asymmetries are of the same order in the
 non-supersymmetric and in the 
supersymmetric case.

\begin{figure}[t]
    \mbox{ }\hfill
    \epsfig{file=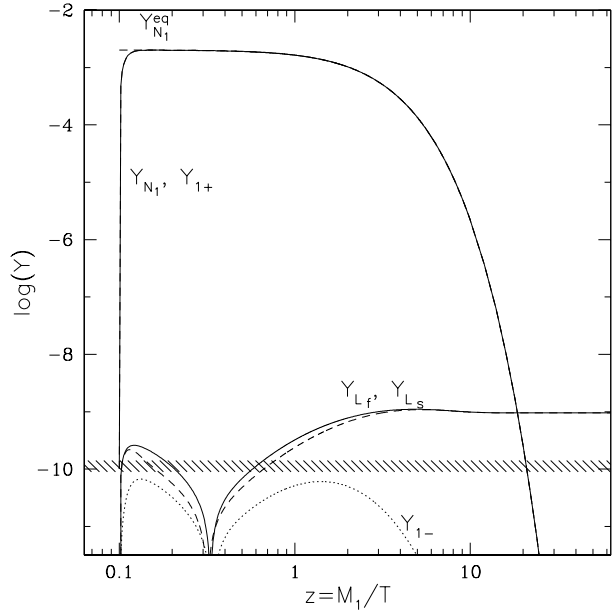,width=8.2cm}
    \hfill\mbox{ }
    \caption{\it Solution of the supersymmetric Boltzmann equations in
      the case of the
     $SU(3)_c\times SU(3)_L\times SU(3)_R\times$ $U(1)_F$ symmetry. 
     \label{asyLR_susy}}
  \end{figure}

Like in the non-supersymmetric scenario it is interesting to see 
whether a parameter choice which gives light neutrino masses and
mixing angles compatible with experimental results, can also explain 
the baryon asymmetry. To this end we again considered the two 
scenarios discussed in sections \ref{su5masses} and \ref{nu_end}. 
The time evolution of the (s)neutrino number densities and the 
(s)lepton asymmetries in the case of the $SU(5)\times U(1)_F$ symmetry
is shown in fig.~\ref{asyB_susy}. Although the CP-asymmetry is now
larger than eq.~(\ref{su5epsilon}) by a factor of two, the wash-out 
processes are more effective than in the non-supersymmetric case, 
and the final asymmetry is again of order $10^{-9}$, corresponding 
to a washout factor $\kappa\sim 0.03$. Similarly, in the case of the 
$SU(3)_c\times SU(3)_L\times SU(3)_R\times U(1)_F$ symmetry 
(cf.~fig.~\ref{asyLR_susy}) the final asymmetry is again of order 
$10^{-9}$. The baryogenesis temperature $T_B \sim 10^{10}$~GeV in the
$SU(5)\times U(1)_F$ model is consistent with the constraint from the
allowed gravitino abundance \cite{bbp98}.

Leptogenesis can also be considered in extended models which contain
heavy $SU(2)$-triplet Higgs fields in addition to right-handed 
neutrinos \cite{ms98,ls98}. Decays of the heavy scalar bosons can in principle
also contribute to the baryon asymmetry. However, since these Higgs particles
carry gauge quantum numbers they are strongly coupled to the plasma and
it is difficult to satisfy the out-of-equilibrium condition. The resulting
large baryogenesis temperature is in conflict with the 
`gravitino constraint' \cite{ds99}.

\subsection{Mass scales of leptogenesis and $B-L$ breaking}

In the previous section we have considered baryogenesis in two models
with hierarchical neutrino masses but rather different flavour structure.
In both cases the Majorana masses of the right-handed neutrinos turned out
to be rather large, i.e., $M_i \geq 10^{10}$~GeV, which implies large mass
scales for the leptogenesis temperature and $B-L$ breaking.  

In order to understand the model dependence of this result it is instructive
to study the case of two generations for which bounds on light and heavy
neutrino masses can be easily obtained (cf.~\cite{by93}).

Without loss of generality the heavy Majorana neutrino mass matrix can 
always be assumed to be diagonal with positive real eigenvalues,
\beq
M = \left(\begin{array}{cc}
    M_1        & 0  \\
    0        & M_2   
    \end{array}\right) \;.
\eeq
Similarly, in terms of the matrix $\wh{m}_D = im_D V_\n$ (cf.~(\ref{seesaw})) 
the light Majorana neutrino mass matrix reads
\beq
m_\n = \wh{m}_D^T{1\over M}\wh{m}_D = \left(\begin{array}{cc}
    m_1        & 0  \\
    0        & m_2   
    \end{array}\right) \;,
\eeq
where the eigenvalues $m_i$ are again real. With
\beq
\wh{m}_D = \left(\begin{array}{cc}
    a        & b  \\
    c        & d   
    \end{array}\right) \;,\quad b=\sqrt{{M_1\over M_2}}\eta d\;,
\quad c=-\sqrt{{M_2\over M_1}}\eta a \;,
\eeq
one obtains
\beq
m_\n = \left(\begin{array}{cc}
    {a^2\over M_1}        & 0  \\
    0        & {d^2\over M_2}  \end{array}\right) (1+\eta^2) \;.
\eeq
Note that $a\ldots d$ are complex. Since $m_1$ and $m_2$ are real there
is only one independent phase. A complete set of parameters for the
two neutrino mass matrices is given by $m_1, m_2, M_1, M_2$ and 
$\eta^2=\r e^{i\a}$.

The out-of-equilibrium condition for the decay of the heavy neutrino
implies an interesting upper bound on the mass of the lightest neutrino
$m_- = \mbox{min} \{ m_1, m_2 \}$. From $\G_{D_1} < H|_{T=M_1}$ one 
obtains for the effective neutrino mass $\wt m_1$,
\beq
\wt m_1 = {(\wh m_D\wt m_D^\dg)_{11}\over M_1} < m_0 = 10^{-3}\ \mbox{eV}\;.
\eeq
With
\beq
\wt m_1 = {m_1 + m_2 |\eta|^2\over |1+\eta^2|}\;,
\eeq
one is led to
\beqa
m_- &<& m_- {1 + |\eta|^2\over |1+\eta^2|} < \wt m_1 \NO\\
&<& 10^{-3}\ \mbox{eV}\;.\label{ubound}
\eeqa
Note, that this bound is only accurate up to a factor ${\cal O}(1)$, as
the values for $\wt m_1$ show which we have obtained for the two models 
discussed above. This difference illustrates
the accuracy to which the out-of-equilibrium condition holds.

The upper bound (\ref{ubound}) can always be satisfied by making the 
neutrino Yukawa couplings very small. However, leptogenesis also requires 
a sufficient amount of CP asymmetry. From (\ref{eps}) one obtains
($M_1 < M_2$),
\beqa
|\ve_1| &\simeq& {3\over 16\p v_2^2} 
{1\over \left(\wh m_D \wh m_D^\dg\right)_{11}}
\mid \mbox{Im}{\left(\wh m_D \wh m_D^\dg\right)_{12}^2}\mid {M_1\over M_2}\NO\\
&=& {3\over 16\p} {|m_2^2-m_1^2|\over v_2^2} {M_1\over m_1 + m_2 |\eta|^2}
{\mid \mbox{Im}{\eta^2}\mid \over |1+\eta^2|}\;.
\eeqa
We now assume a maximal phase, i.e., $\eta^2 = i\r$, $\a=\p/2$, and
$|\ve_1| > \ve_0 \sim 10^{-8}$, which is the smallest value required by
the observed baryon asymmetry $Y_B \sim 10^{-10}$. This yields for the
heavy neutrino mass $M_1$,
\beq
M_1 > {16\p\over 3} {v_2^2\over |m_2^2-m_1^2|} \ve_0 
{(1+\r^2)^{1/2}(m_1+m_2\r)\over \r}\;.
\eeq
Since the right-hand side has a minimum at $\r = (m_1/m_2)^{1/3}$,
one obtains a lower bound on $M_1$,
\beq\label{lbound}
M_1 > {16\p\over 3} \ve_0 v_2^2 
{\left(m_1^{2/3}+m_2^{2/3}\right)^{3/2}\over |m_2^2-m_1^2|}\;.
\eeq
For quasi-degenerate neutrinos the bound (\ref{ubound}) implies
$m_1 \sim m_2 < m_0 \sim 10^{-3}\ \mbox{eV}$. This yields
\beq\label{lM1}
M_1 > 10^{18}\ \mbox{GeV}\ \ve_0 \sim 10^{10}\ \mbox{GeV}\;,
\eeq
independent of our knowledge about the solar and atmospheric neutrino
anomalies.
For hierarchical neutrinos all masses have to satisfy the upper bound on
the electron neutrino mass obtained from $\b$-decays since the mass
differences infered from the athmospheric and solar neutrino anomalies
are much smaller. Hence, one has $m_- \ll m_+ <  1 \mbox{eV}$ which implies
\beq\label{lM2}
M_1 > 10^{15}\ \mbox{GeV}\ \ve_0 \sim 10^7\ \mbox{GeV}\;.
\eeq
The bounds (\ref{lM1}) and (\ref{lM2}) confirm the expectation, in accord
with the explicit models discussed above, that the smallness of the light
neutrino masses suggested by experimental data as well as the 
out-of-equilibrium condition requires heavy Majorana neutrinos
far above the electroweak scale. The bounds (\ref{lM1}) and (\ref{lM2})
can be relaxed in the case $|M_1-M_2|={\cal O}(\G_i)$. Then a resonant
enhancement of the CP asymmetry can occur \cite{pil99}. In this case the
lower bound $M_1 > 10^5$~GeV has been obtained in a recent analysis 
\cite{flo99}.

In the context of a grand unified theory the large Majorana masses will
be generated by a Higgs mechanism which breaks $B-L$ spontaneously. 
The corresponding vector bosons have to be sufficiently heavy such
that the out-of-equilibrium condition for the decaying Majorana neutrino
$N_1$ is not violated by processes like
$N_1 N_1 \rightarrow Z' \rightarrow e e^c$ \cite{lut92}. Assuming Higgs 
couplings $h_r < 1$ and a gauge coupling ${\cal O}(1)$ one obtains the
lower bound on the scale of $B-L$ breaking
\beq
\L_{B-L} \sim M_{Z'} 
> 10^{11} \left({M_1\over 10^{10}\mbox{GeV}}\right)^{3/4}\;.
\eeq
In the case of hierarchical heavy Majorana neutrinos the scale of 
$B-L$ breaking is naturally identified with the grand unification scale,
as the models illustrate which we discussed in the previous section.

\section{Outlook}

Detailed studies of the thermodynamics of the electroweak interactions at
high temperatures have shown that in the standard model and most of its
extensions the electroweak transition is too weak to affect the 
cosmological baryon asymmetry. Hence, one has to search for baryogenesis
mechanisms above the Fermi scale. 

Due to sphaleron processes baryon number and lepton number are related
in the high-temperature symmetric phase of the standard model. As a
consequence, the cosmological baryon asymmetry is related to neutrino
properties. Generically, baryogenesis requires lepton number violation, which 
occurs in many extensions of the standard model with right-handed neutrinos and
Majorana neutrino masses. In detail the relations between $B$, $L$ and $B-L$
depend on all other processes taking place in the plasma, and therefore
also on the temperature.  

Although lepton number violation is needed in order to obtain a baryon
asymmetry, it must not be too strong since otherwise any baryon and lepton
asymmetry would be washed out. Hence, leptogenesis leads to stringent upper 
and lower bounds on the masses of the light and heavy Majorana neutrinos,
respectively.

The solar and atmospheric neutrino deficits can be interpreted as a result
of neutrino oscillations. For hierarchical neutrinos the corresponding
neutrino masses are very small. Assuming the see-saw mechanism, this suggests
the existence of very heavy right-handed neutrinos and a large scale of
$B-L$ breaking.

It is remarkable that these hints on the nature of lepton number
violation fit very well together with the idea of leptogenesis.  For
hierarchical neutrino masses, with $B-L$ broken at the unification
scale $\Lambda_{\mbox{\scriptsize GUT}}\sim 10^{16}\;$GeV, the
observed baryon asymmetry $n_B/s \sim 10^{-10}$ is naturally explained
by the decay of heavy Majorana neutrinos. The corresponding
baryogenesis temperature is $T_B \sim 10^{10}$ GeV.

The cosmological baryon asymmetry provides important constraints on 
neutrino masses and mixings. In addition the connection between 
lepton flavour and quark flavour changing processes can be studied in 
unified theories. In supersymmetric models implications for the mass spectrum 
of superparticles can be derived from the cosmological bound on the
gravitino number density. It is intriguing that the baryogenesis temperature 
$T_B$ is of the same order as the supersymmetry breaking scale $\L_S$.
It will be interesting to see to what extent further theoretical work
and experimental data will be able to identify or exclude leptogenesis
as the dominant source of the cosmological matter-antimatter
asymmetry. In order to obtain
a theory of leptogenesis one also has to go beyond the classical Boltzmann
equations and to treat the generation of the baryon asymmetry fully 
quantum mechanically, which presents a challenging problem of non-equilibrium
physics.

\end{document}